\documentstyle[aps,epsf,eqsecnum,preprint]{revtex}
\tightenlines
\begin{document}
\draft \preprint{{\bf Published in Phys. Rev. D 62, 056011
(2000)}}
\title{Many pion decays of $\rho(770)$ and $\omega(782)$ mesons
in  chiral theory.}
\author{N.~N.~Achasov \footnote{Electronic address: achasov@math.nsc.ru}
and
A.~A.~Kozhevnikov \footnote{Electronic address: kozhev@math.nsc.ru}}
\address{Laboratory of Theoretical Physics, \\
Sobolev Institute for Mathematics \\
630090, Novosibirsk-90, Russia}
\date{\today}
\maketitle
\begin{abstract}
The decays  $\rho(770)\to4\pi$ and $\omega(782)\to5\pi$ are
considered in detail in the approach based on the Weinberg
Lagrangian obtained upon the nonlinear realization of chiral
symmetry, added with the term induced by the anomalous Lagrangian
of Wess and Zumino. The partial widths and excitation curves of
the decays $\rho^0\to2\pi^+2\pi^-$, $\pi^+\pi^-2\pi^0$,
$\rho^\pm\to2\pi^\pm\pi^\mp\pi^0$, $\rho^\pm\to\pi^\pm3\pi^0$ are
evaluated for $e^+e^-$ annihilation,  photoproduction and $\tau$
lepton decays. The results of calculations are compared with the
recent CMD-2 data on the decay $\rho^0\to2\pi^+2\pi^-$ observed in
$e^+e^-$ annihilation. The $\omega\to5\pi$ decay widths and
excitation curves in $e^+e^-$ annihilation are obtained. The
angular distributions for various combinations of the final pions
in the decays $\rho\to4\pi$ and $\omega\to5\pi$ are written. The
perspectives of the experimental study of the above decays  in
$e^+e^-$ annihilation, $\tau$ lepton decays and photoproduction
are discussed.
\end{abstract}
\narrowtext

\section{Introduction}
\label{sec1}

Despite the lack of a straightforward derivation from  first
principles of QCD, the effective Lagrangians that describe the low-energy
interactions of the ground state octet of pseudoscalar
mesons $\pi$, $K$, $\eta$ are constructed upon treating these
mesons as the Goldstone bosons of the spontaneously broken chiral
$U_L(3)\times U_R(3)$ symmetry of the massless three-flavored QCD
Lagrangian. The key point in this task is that the transformation
properties of the Goldstone fields under the nonlinear realization
of chiral symmetry are sufficient for the establishing the most
general form of the effective Lagrangian \cite{weinberg79}. As far
as vector mesons are concerned, the situation is not so clear,
because the vector mesons, contrary to the pseudoscalar ones,
cannot be considered as the Goldstone bosons of the spontaneously
broken symmetry.  For this reason there exist different schemes of
including these mesons into the effective chiral Lagrangians.  The
scheme of Ref.~\cite{bando} where the vector mesons are treated as
the dynamical gauge bosons of hidden local symmetry (HLS),
incorporates these mesons into the effective chiral Lagrangian in
a most elegant way.  However, a straightforward comparison
with the data of the predictions of these models immediately meets
with  difficulty, since  the dominant decay modes of all the ground
state vector mesons produce  pseudoscalar mesons that are far
from being soft. This means that both the higher derivatives in
the low energy expansion of the effective Lagrangians and the loop
quantum corrections should be taken into account.

We suggest here   many pion decay modes of  $\rho(770)$ and
$\omega(782)$ mesons as a test ground for the chiral models of
vector mesons \cite{ach99,ach00} . Theoretically, pions in the
final state in such decays are soft to the extent to be specified
below. This fact allows one to neglect the higher derivative and
loop corrections in the effective Lagrangian. On the other hand,
the unconventional, from the point of view of the chiral pion
dynamics, sources of soft pions are feasible. Indeed, the progress
in increasing the intensity of low energy $e^+e^-$ colliders
($\phi$ factories) \cite{phifac}, photon beams \cite{cebaf}, and a
huge number of the specific hadronic decays of $\tau$ leptons
could offer the naturally controlled sources of soft pions,
provided the sufficiently low invariant mass regions of hadronic
systems are isolated. The yield of pions is considerably enhanced
when they are produced through the proper vector resonance states,
which will hopefully offer the possibility of testing the
mentioned models.

Pions emitted in the  decay  $\rho\to4\pi$ are rather soft,
because their typical momentum is $|{\bf p}|\sim m_\pi$, where
$m_\pi$ is the pion mass. By this reason this decay attracts much
attention \cite{rittenberg69,bramon93,kuraev95,birse96} from the
point of view of the study of the predictions of the effective
chiral Lagrangians for vector mesons. As was found in Refs.
\cite{rittenberg69,bramon93,kuraev95}, the above decay should be
rather strong, B$(\rho\to4\pi)\sim10^{-4}$. The calculations of
Refs. \cite{bramon93,kuraev95} were analyzed in detail in Ref.
\cite{birse96}, where a number of shortcomings of the former
related with the actual violation of chiral invariance, in
particular, the Adler condition \cite{adler} for soft pions, was
uncovered. The correct results based on the amplitudes obeying the
Adler condition and obtained in Refs. \cite{ach99,birse96}, correspond
to B$(\rho\to4\pi)\approx10^{-5}$. The large magnitude of the
branching ratio B$(\rho\to4\pi)\sim10^{-4}$ obtained in
Ref. \cite{rittenberg69} is related, in all appearance, with a very
rough method of calculation. A common drawback of  Refs.
\cite{rittenberg69,bramon93,kuraev95,birse96} is that their
authors evaluate the partial width  at the only energy equal to
the mass of the $\rho$, as if the latter were a genuine narrow
peak. However, the fact that the width of the $\rho$ resonance is
rather large, and $\Gamma(\rho\to4\pi,E)$ rises rapidly with the
energy  increase even at energies inside the $\rho$ peak, forces
one to think that the magnitude of the $4\pi$ partial width at the
$\rho$ mass cannot be an adequate characteristics of the dynamics
of the process. In this respect, it is just the resonance
excitation curve in the channel $e^+e^-\to \rho^0\to4\pi$ which is
of much interest, being a test ground of various chiral models of
the decay under consideration.

Pions emitted in the decay $\omega\to5\pi$ \cite{ach00} are truly
soft, because they possess the typical momentum $|{\bf
p}|\simeq0.5m_\pi$. Hence, the lowest order terms obtained upon
neglecting  the higher derivatives and loop corrections should
give the reliable results.

The aim of the present paper is to consider in detail the many
pion decay modes of the vector $\rho(770)$ and $\omega(782)$
mesons in the framework of the effective chiral Lagrangian
approach. The first Lagrangian of this kind incorporating the
isotopic triplets of the $\rho$ meson field $\bbox{\rho}_\mu$, the
pion field $\bbox{\pi}$, and their interaction was proposed by
Weinberg \cite{weinberg68} under the nonlinear realization of the
chiral symmetry
\begin{eqnarray}
{\cal
L}^{(\rho+\pi)}&=&-{1\over4}\left(\partial_\mu\bbox{\rho}_\nu-
\partial_\nu\bbox{\rho}_\mu+g[\bbox{\rho}_\mu\times\bbox{\rho}_\nu]
\right)^2     \nonumber\\
&&+{m^2_\rho\over2}\left[\bbox{\rho}_\mu+
{[\bbox{\pi}\times\partial_\mu\bbox{\pi}]\over2g f^2_\pi
(1+\bbox{\pi}^2/4f^2_\pi)}\right]^2      \nonumber\\
&&+{(\partial_\mu\bbox{\pi})^2\over2\left(1+\bbox{\pi}^2
/4f^2_\pi\right)^2}
-{m^2_\pi\bbox{\pi}^2\over2(1+\bbox{\pi}^2/4f^2_\pi)},
\label{lwein}
\end{eqnarray}
where $f_\pi=92.4$ MeV is the pion decay constant, and the cross
denotes the vector product in the isovector space. As was shown,
the Lagrangian Eq.~(\ref{lwein}) results from the HLS approach
\cite{bando}. The $\rho\rho\rho$ coupling constant $g$ and the
$\rho\pi\pi$ coupling constant $g_{\rho\pi\pi}$ are related to the
$\rho$ mass $m_\rho$ and the pion decay constant $f_\pi$ via the
parameter of hidden local symmetry $a$ as \cite{bando}
\begin{eqnarray}
g&=&m_\rho/f_\pi\sqrt{a},   \nonumber\\
g_{\rho\pi\pi}&=&\sqrt{a}m_\rho/2f_\pi. \label{a}
\end{eqnarray}
Note that  $a=2$, if one demands the universality condition
$g=g_{\rho\pi\pi}$ to be satisfied. Then the so called
Kawarabayashi-Suzuki-Riazzuddin-Fayyazuddin (KSRF) relation
\cite{ksrf} arises
\begin{equation}
2g^2_{\rho\pi\pi}f^2_\pi/m^2_\rho=1 \label{ksrf}
\end{equation}
which beautifully agrees with experiment. The $\rho\pi\pi$
coupling constant resulting from this relation is
$g_{\rho\pi\pi}=5.89$. The Lagrangian of Eq.~(\ref{lwein}) should
be added with the kinetic and mass terms of the  isosinglet
$\omega(782)$ field $\omega_\mu$,
\begin{equation}
{\cal
L}^{(\omega)}=-{1\over4}(\partial_\mu\omega_\nu-\partial_\nu\omega_\mu)^2
+{m^2_\omega\over2}\omega^2_\mu, \label{lomega}
\end{equation}
and the term describing the interaction of $\omega$ with
$\bbox{\rho}$ and $\bbox{\pi}$. The latter comes from the term
induced by the anomalous Lagrangian of Wess and Zumino
\cite{bando,wza},
\begin{equation}
{\cal L}^{(\omega\rho\pi)}={N_cg^2\over8\pi^2 f_\pi}
\varepsilon_{\mu\nu\lambda\sigma}\partial_\mu\omega_\nu\left(\bbox{\pi}
\cdot\partial_\lambda\bbox{\rho}_\sigma\right), \label{wz}
\end{equation}
where $N_c=3$ is the number of colors, and
$\varepsilon_{\mu\nu\lambda\sigma}$ is the antisymmetric unit
tensor.  In agreement with Ref. \cite{bando}, the contribution of the
pointlike vertex $\omega\to3\pi$ is omitted. See, however, the
discussion following Eq.~(\ref{b3pi0}) in Sec.~\ref{sec4} where
the role of this vertex in the $\omega\to5\pi$ decay is briefly
discussed.

The HLS approach \cite{bando} permits one to include the axial
mesons as well \cite{fn1}. An ideal treatment would consist of
following the line of reasoning that under the assumption of
$m_\rho\sim E\ll m_{a_1}$, the difference between the models with
and without $a_1$ meson were reduced to the taking into account
the higher derivatives \cite{fn2} expected to be small. In the
real life, one has $m^2_{a_1}-m^2_\rho\sim m^2_\rho$, and the
correction may appear to be appreciable even at the $\rho$ mass.
In fact, the calculation \cite{birse96} shows that the corrections
amount to $\sim20\mbox{ - }30\%$ in the width. This means, in
particular, that the left shoulder of the $\rho$ peak, where the
contributions of higher derivatives are vanishing rapidly, is the
best place to work. In the present paper, we do not take into
account the $a_1$ meson contribution in the $\rho\to4\pi$ decay
amplitude. In the meantime, such a contribution is negligible, as
will be clear later on, in the $\omega\to5\pi$ decay amplitude.

The rest of the paper is devoted to the working out the
consequences of the above Lagrangians for the many pion decays of
the $\rho(770)$ and $\omega(782)$ mesons. Specifically,  the
partial widths and resonance excitation curves are calculated for
the reactions $e^+e^-\to\rho^0\to2\pi^+2\pi^-$ and
$e^+e^-\to\rho^0\to\pi^+\pi^-2\pi^0$. It is shown that the
intensities of the above decays change as fast as two times the
phase space variation, upon the energy variation inside the $\rho$
widths. All this means that $e^+e^-$ offers an ideal tool for the
study of such effects. The decay widths of charged $\rho$ meson,
$\rho^\pm\to\pi^\pm3\pi^0$ and  $\rho^\pm\to2\pi^\pm\pi^\mp\pi^0$,
as well as $\omega$ meson, $\omega\to2\pi^+2\pi^-\pi^0$ ¨
$\omega\to\pi^+\pi^-3\pi^0$, are also evaluated. We choose these
particular modes because the final pions in the decay
$\rho\to4\pi$ are practically soft while in the decay
$\omega\to5\pi$ they are soft. Note that the final pions in the
G-parity violating decay $\rho\to3\pi$ are not sufficiently soft
to compare the calculated branching ratio with the predictions of
chiral models.  The final pions in the decay $\omega\to4\pi$ are
sufficiently soft to make such a comparison meaningful. However,
the branching ratios of both above G-parity violating decays
$\rho\to3\pi$ and $\omega\to4\pi$ are determined completely by the
$\omega\rho$ transition amplitude and by the well known decay
$\omega\to3\pi$ and the decay $\rho\to4\pi$, respectively.

The following material  is organized as follows. Section
\ref{sec2} contains the expressions for the $\rho\to4\pi$
amplitudes. The results of calculation of the  excitation curves
and partial widths for different isotopic states of four pions are
presented in Sec.~\ref{sec3}. This task is fulfilled in the cases
of $e^+e^-$ annihilation, $\tau$ decays, and photoproduction. The
partial widths of the decays $\omega\to5\pi$ are discussed in
Sec.~\ref{sec4}. Although  an appreciable part of the material of
Sec.~\ref{sec4} is contained in Ref. \cite{ach00}, we include here
the basic results from that paper in order to keep the
presentation self-contained.  Section \ref{sec5} contains concluding
remarks. The angular distributions of the various combinations of
emitted pions in the decays $\rho\to4\pi$ and $\omega\to5\pi$
obtained for the case of $e^+e^-$ annihilation are presented in
the Appendix.

\section{Amplitudes of $\rho\to4\pi$ decay}
\label{sec2}

\begin{figure}
\centerline {\epsfysize=7in \epsfbox{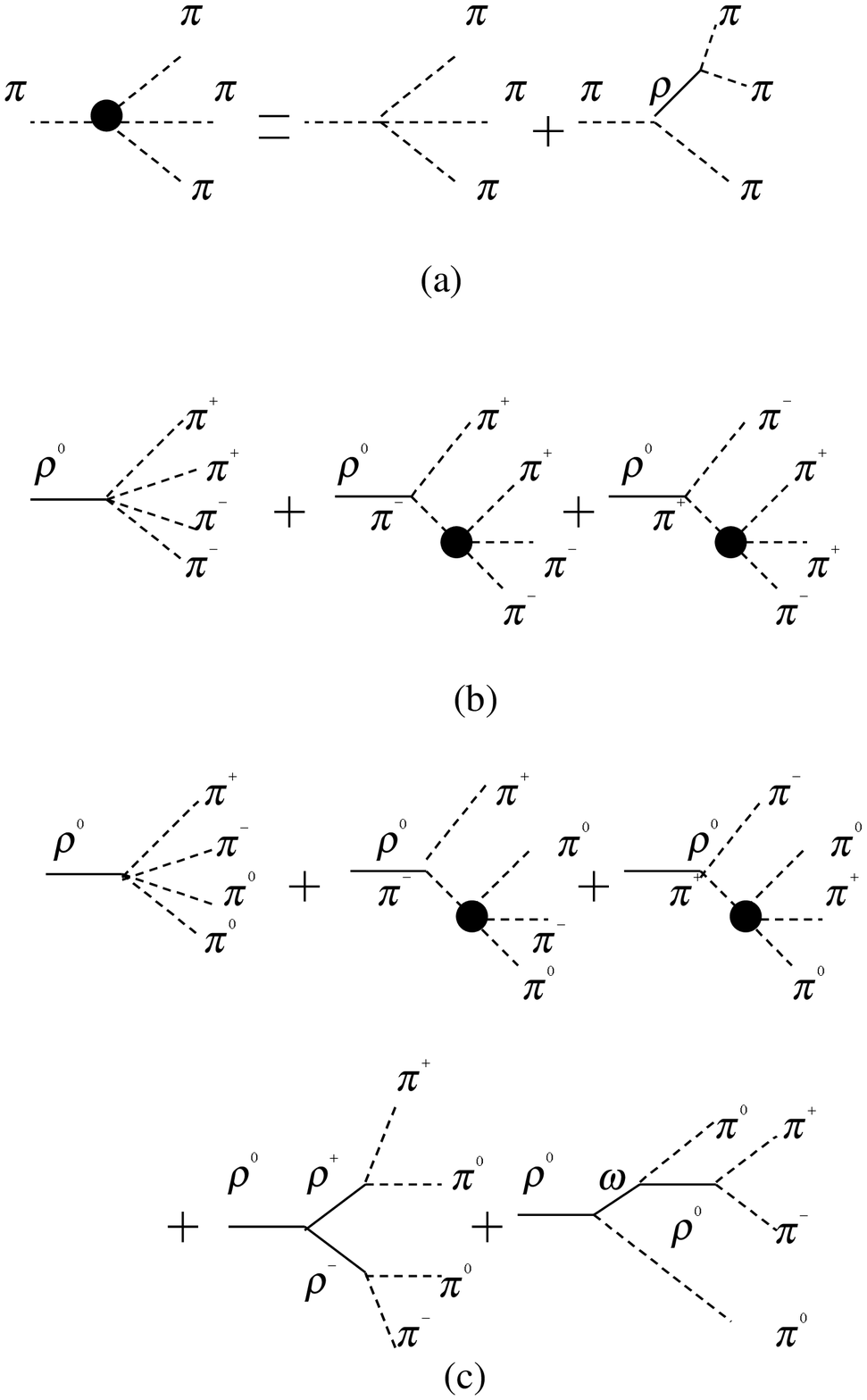}} \caption{(a) The
diagrams describing the $\pi\to3\pi$ transition amplitude. The
symmetrization over  momenta of identical pions is understood when
necessary. The diagrams describing the amplitudes of the decays
(b) $\rho^0\to\pi^+\pi^-\pi^+\pi^-$  and (c)
$\rho^0\to\pi^+\pi^-\pi^0\pi^0$. The shaded circles in the
$\pi\to3\pi$ vertices  in the diagrams (b) and (c) refer to the
sum of diagrams shown in (a). The symmetrization over  momenta of
identical pions emitted from different vertices is implied. The
diagrams for the decays $\rho^+\to\pi^+\pi^0\pi^0\pi^0$ and
$\rho^+\to\pi^+\pi^+\pi^-\pi^0$ are analogous to those of (b) and
(c), respectively. \label{fig1}}
\end{figure}

The amplitudes of the decays of our interest are obtained from the
Weinberg Lagrangian Eq.~(\ref{lwein}). First, let us obtain the
$\pi\to3\pi$ transition amplitudes necessary for the calculation
of the many-pion decays of vector mesons. They are given by the
diagrams shown in Fig.~\ref{fig1}(a) and look as \widetext
\begin{eqnarray}
M(\pi^+\to\pi^+_{q_1}\pi^+_{q_2}\pi^-_{q_3})&=&(1+P_{12}){1\over2f^2_\pi}
\left\{-2(q_1,q_2)+a(q_1,q_2-q_3)\left[1-{m^2_\rho\over D_\rho(q_2+q_3)}
\right]\right\},
\nonumber\\
M(\pi^+\to\pi^+_{q_1}\pi^0_{q_2}\pi^0_{q_3})&=&(1+P_{23}){1\over2f^2_\pi}
\left\{(q_3,q_1-2q_2)+
a(q_3,q_1-q_2)\left[1-{m^2_\rho\over D_\rho(q_1+q_2)}\right]\right\},
\nonumber\\
M(\pi^0\to\pi^+_{q_1}\pi^-_{q_2}\pi^0_{q_3})&=&(1+P_{12}){1\over2f^2_\pi}
\left\{(q_1,q_2-2q_3)-
a(q_1,q_2-q_3)\left[1-{m^2_\rho\over D_\rho(q_2+q_3)}\right]\right\},
\nonumber\\
M(\pi^0\to\pi^0_{q_1}\pi^0_{q_2}\pi^0_{q_3})&=&-{1\over f^2_\pi}
\left[(q_1,q_2)+(q_1,q_3)+(q_2,q_3)\right].
\label{4pi}
\end{eqnarray}
\narrowtext
where $P_{ij}$ stands for the operator of the interchange of the pion momenta
$q_i$ and $q_j$, and
\begin{equation}
D_\rho(q)=m^2_\rho-q^2-im^2_\rho\left({q^2-4m_\pi^2\over m^2_\rho-4m^2_\pi}
\right)^{3/2}{\Gamma_{\rho\pi\pi}(m^2_\rho)\over\sqrt{q^2}}
\label{proprho}
\end{equation}
is the inverse propagator of $\rho$ meson.
Our notation for the Lorentz invariant scalar product of two four vectors
$a$ and $b$ hereafter is $(a,b)=a_0b_0-{\bf a}\cdot{\bf b}$.
As it will be clear later on, the nonrelativistic expressions for the above
amplitudes are needed. They are obtained upon neglecting the space components
of the pion four momenta and look as
\begin{eqnarray}
M(\pi^+\to\pi^+_{q_1}\pi^+_{q_2}\pi^-_{q_3})&=&-{2m^2_\pi\over f^2_\pi},
\nonumber\\
M(\pi^+\to\pi^+_{q_1}\pi^0_{q_2}\pi^0_{q_3})&=&-{m^2_\pi\over f^2_\pi},
\nonumber\\
M(\pi^0\to\pi^+_{q_1}\pi^-_{q_2}\pi^0_{q_3})&=&-{m^2_\pi\over f^2_\pi},
\nonumber\\
M(\pi^0\to\pi^0_{q_1}\pi^0_{q_2}\pi^0_{q_3})&=&-{3m^2_\pi\over f^2_\pi}.
\label{4pinr}
\end{eqnarray}
Note that the HLS parameter $a$ drops from the expressions in the
nonrelativistic limit.

The diagrams representing  the amplitudes of the decays
$\rho\to4\pi$ for different combinations of the charges of final
pions are shown in Fig.~\ref{fig1}(b) and (c). Introducing the
four vector of polarization of the decaying $\rho$ meson
$\varepsilon_\mu$ one can write the general expression for the
amplitude in the form $$M={g_{\rho\pi\pi}\over
f^2_\pi}\varepsilon_\mu J_\mu,$$ where
\begin{equation}
g_{\rho\pi\pi}/f^2_\pi=\sqrt{a}m_\rho/2f^3_\pi \label{a1}
\end{equation}
results from Eq.~(\ref{a}).
Let us give the expressions for the current $J_\mu$ for all the decay
modes considered here.\\
(1) The decay $\rho^0(q)\to\pi^+(q_1)\pi^+(q_2)\pi^-(q_3)\pi^-(q_4)$.
One has
\widetext
\begin{eqnarray}
J_\mu&=&(1+P_{12})(1+P_{34})\left\{-q_{1\mu}\left[{1\over2}
+{a(q_2,q_3)-(a-2)(q_3,q_4)\over D_\pi(q-q_1)}\right]
\right.     \nonumber\\
&&\left.+q_{3\mu}\left[{1\over2}
+{a(q_1,q_4)-(a-2)(q_1,q_2)\over D_\pi(q-q_3)}\right]\right.   \nonumber\\
&&\left.+am^2_\rho(1+P_{13})
{q_{1\mu}(q_3,q_2-q_4)\over D_\pi(q-q_1)D_\rho(q_2+q_4)}\right\}.
\label{eech}
\end{eqnarray}
\narrowtext Hereafter  $D_\pi(q)=m^2_\pi-q^2$   is the inverse
propagator of pion.\\
(2) The decay
$\rho^0(q)\to\pi^+(q_1)\pi^-(q_2)\pi^0(q_3)\pi^0(q_4)$. In this
case one has $J_\mu=J^{\rm nan}_\mu+J^{\rm an}_\mu$, where
\widetext
\begin{eqnarray}
J^{\rm nan}_\mu&=&-(1-P_{12})(1+P_{34})q_{1\mu}\left\{{1\over4}+
{1\over D_\pi(q-q_1)}\left[(a-1)(q_3,q_4)-(a-2)(q_2,q_3)
\right.\right.\nonumber\\
&&\left.\left.+am^2_\rho
{(q_3,q_2-q_4)\over D_\rho(q_2+q_4)}\right]\right\}  \nonumber\\
&&+(1+P_{34}){m^2_\rho\over2 D_\rho(q_1+q_3)D_\rho(q_2+q_4)}
\left[(q_1+q_3-q_2-q_4)_\mu(q_1-q_3,q_2-q_4)\right.\nonumber\\
&&\left.-2(q_1-q_3)_\mu(q_1+q_3,q_2-q_4)+
2(q_2-q_4)_\mu(q_2+q_4,q_1-q_3)\right]
\label{eenena}
\end{eqnarray}
\narrowtext is obtained from Eq. (\ref{lwein}), while the
contribution of the term induced by the anomalous Lagrangian of
Wess and Zumino \cite{bando,wza} manifesting in the process
$\rho^0\to\omega\pi^0\to \pi^+\pi^-\pi^0\pi^0$, is given by the
expression \widetext
\begin{eqnarray}
J^{\rm an}_\mu&=&2\left({N_cg^2\over8\pi^2}\right)^2
(1+P_{34})\left[q_{1\mu}(1-P_{23})(q,q_2)(q_3,q_4)\right.\nonumber\\
&&\left.+q_{2\mu}(1-P_{13})(q,q_3)(q_1,q_4)+
q_{3\mu}(1-P_{12})(q,q_1)(q_2,q_4)\right]        \nonumber\\
&&\times\left[{1\over D_\rho(q_1+q_2)}+{1\over D_\rho(q_1+q_3)}
+{1\over D_\rho(q_2+q_3)}\right]{1\over D_\omega(q-q_4)},
\label{eenean}
\end{eqnarray}
\narrowtext where $D_\omega(q)=m^2_\omega-q^2$ is the inverse
$\omega$ meson propagator.  In general, this term is attributed to
the contribution of higher derivatives. Nevertheless, we take it
into account to show the effect of the latter and the dynamical
effect of the opening of the channel $\rho\to\omega\pi\to4\pi$.
The following amplitudes of the charged $\rho$ decay are necessary
for obtaining the $\omega\to5\pi$ decay amplitude, and are of
their own interest when studying the reactions of peripheral
$\rho$ meson production and $\tau$ decays.\\ (3) The decay
$\rho^+(q)\to\pi^+(q_1)\pi^0(q_2)\pi^0(q_3)\pi^0(q_4)$. One has
\widetext
\begin{eqnarray}
J_\mu&=&(1+P_{24}+P_{34})\left\{2q_{1\mu}\left[{1\over3}+
{(q_2,q_3)\over D_\pi(q-q_1)}\right]-
{q_{4\mu}\over D_\pi(q-q_4)}\left[2(a-1)(q_2,q_3)\right.\right.
\nonumber\\
&&\left.\left.-(a-2)(q_1,q_2+q_3)
+am^2_\rho(1+P_{23}){(q_2,q_1-q_3)\over D_\rho(q_1+q_3)}\right]\right\}.
\label{peneu}
\end{eqnarray}
\narrowtext (4) The decay
$\rho^+(q)\to\pi^+(q_1)\pi^+(q_2)\pi^-(q_3)\pi^0(q_4)$. Here, the
contribution induced by the anomalous Lagrangian of Wess and
Zumino  is also possible, hence $J_\mu=J^{\rm nan}_\mu+J^{\rm
an}_\mu$,  where \widetext
\begin{eqnarray}
J^{\rm nan}_\mu&=&(1+P_{12})\left\{{1\over2}(q_1-q_4)\mu+
{q_{1\mu}(1+P_{23})\over D_\pi(q-q_1)}\left[(a-1)(q_2,q_3)-(a-2)(q_2,q_4)
\right]\right.        \nonumber\\
&&\left.-
{q_{4\mu}\over D_\pi(q-q_4)}\left[a(q_1,q_3)-(a-2)(q_1,q_2)\right]
\right.      \nonumber\\
&&\left.-am^2_\rho\left[{q_{1\mu}\over D_\pi(q-q_1)}(1+P_{23})
{(q_2,q_3-q_4)\over D_\rho(q_3+q_4)}+{q_{4\mu}(q_1,q_2-q_3)
\over D_\pi(q-q_4)D_\rho(q_2+q_3)}\right]\right.
\nonumber\\
&&\left.+{m^2_\rho\over2D_\rho(q_1+q_3)D_\rho(q_2+q_4)}
\left[(q_1+q_3-q_2-q_4)_\mu(q_1-q_3,q_2-q_4)\right.\right.\nonumber\\
&&\left.\left.-2(q_1-q_3)_\mu(q_1+q_3,q_2-q_4)
+2(q_2-q_4)_\mu(q_1-q_3,q_2+q_4)\right]
\right\}
\label{pechnan}
\end{eqnarray}
\narrowtext
is obtained from Eq.~(\ref{lwein}), while the term induced by the
anomaly looks as
\widetext
\begin{eqnarray}
J^{\rm an}_\mu&=&2\left({N_cg^2\over8\pi^2}\right)^2
(1+P_{23})\left[q_{1\mu}(1-P_{24})(q,q_4)(q_2,q_4)\right.\nonumber\\
&&\left.+q_{2\mu}(1-P_{14})(q,q_1)(q_3,q_4)+
q_{4\mu}(1-P_{12})(q,q_2)(q_1,q_3)\right]        \nonumber\\
&&\times\left[{1\over D_\rho(q_1+q_2)}+{1\over D_\rho(q_1+q_4)}
+{1\over D_\rho(q_2+q_4)}\right]{1\over D_\omega(q-q_3)}.
\label{pechan}
\end{eqnarray}
\narrowtext
One can verify that up to the corrections of the order of
$\sim m^2_\pi/m^2_\rho$, the above written amplitudes vanish in the limit
of vanishing 4-momentum of each final pion. In other words, they
obey the Adler condition.

It is useful to obtain the nonrelativistic expressions for the $\rho\to4\pi$
decay amplitudes which are relevant for the four pion invariant mass below
700 MeV. This can be made upon neglecting the space components of the
pion momenta.  One can convince oneself that $a$ enters the expressions for
amplitudes as an overall factor Eq.~(\ref{a1}) in this limit, so that the
latter look as
\begin{eqnarray}
M(\rho^0\to\pi^+_{q_1}\pi^+_{q_2}\pi^-_{q_3}\pi^-_{q_4})
&\simeq&-{g_{\rho\pi\pi}\over2f^2_\pi}
(\varepsilon,q_1+q_2-q_3-q_4),      \nonumber\\
M(\rho^0\to\pi^+_{q_1}\pi^-_{q_2}\pi^0_{q_3}\pi^0_{q_4})
&\simeq&-{g_{\rho\pi\pi}\over4f^2_\pi} (\varepsilon,q_1-q_2),
\nonumber\\
M(\rho^+\to\pi^+_{q_1}\pi^0_{q_2}\pi^0_{q_3}\pi^0_{q_4})
&\simeq&{g_{\rho\pi\pi}\over f^2_\pi}(\varepsilon,q_1),\nonumber\\
M(\rho^+\to\pi^+_{q_1}\pi^+_{q_2}\pi^-_{q_3}\pi^0_{q_4})
&\simeq&{g_{\rho\pi\pi}\over4f^2_\pi} (\varepsilon,q_1+q_2-2q_4).
\label{nonrel}
\end{eqnarray}
The amplitudes of the four pion decays of the $\rho^-$ are
obtained from corresponding expressions for the $\rho^+$ by
reversing an overall sign. These considerably simplified
expressions are especially convenient in the calculation of the
$\omega\to5\pi$ decay amplitude, because the typical invariant
masses of the four pion system in the above decay are in the
vicinity of 620 MeV (see Sec.~\ref{sec4} for more detail).

\section{Results for various $\rho\to4\pi$ decays}
\label{sec3}

When evaluating the partial widths of the $4\pi$ decays of $\rho$
meson the modulus squared of the matrix element is expressed via
the Kumar variables \cite{kumar}. The idea of speeding up the
numerical integration suggested in Ref.~\cite{sag} is realized in
the numerical algorithm. The results of evaluation of the partial
widths at $\sqrt{s}=m_\rho=770$ MeV are as follows:
$\Gamma(\rho^0\to2\pi^+2\pi^-,m_\rho)=0.89$ keV,
$\Gamma(\rho^0\to\pi^+\pi^-2\pi^0,m_\rho)=0.24$  and 0.44 keV,
respectively, without and with the induced anomaly term being
taken into account. This coincides with the results obtained in
Ref. \cite{birse96}. In the case of the charged $\rho$ meson decays it
is obtained for the first time:
$\Gamma(\rho^+\to\pi^+3\pi^0,m_\rho)=0.41$ keV,
$\Gamma(\rho^+\to2\pi^+\pi^-\pi^0,m_\rho)=0.71$  and 0.90 keV
respectively, without and with the anomaly induced term being
taken into account. When obtaining these figures, the narrow
$\rho$ width approximation is used. This is equivalent to the
setting $\Gamma_{\rho\pi\pi}\to0$ in Eq.~(\ref{proprho}). Keeping
the physical value of the $\rho$ width gives the results deviating
from those obtained in the narrow width approximation by a
quantity that does not exceed a few percents of the values
obtained in the latter. This is true in the case of the invariant
mass of the four pion state lying below the $\rho\pi$ threshold
energy, $m_{4\pi}<910$ MeV.  Recall that the allowing for the
finite widths effects is in fact equivalent to the loop correction
being taken into account.

The above results are obtained at $a=2$. The variation of $a$
within $20\%$ around this value implies the variation of the branching
ratios within $20\%$ around the values cited above. This fact can be easily
traced in the nonrelativistic limit where the parameter $a$
enters the expressions for the amplitudes as an overall factor $\sqrt{a}$,
see Eqs.~(\ref{a}) and (\ref{nonrel}).

\subsection{The decay $\rho^0\to4\pi$ as manifested in $e^+e^-$
annihilation.}
\label{susecee}

\begin{figure}
\centerline {\epsfysize=7in \epsfbox{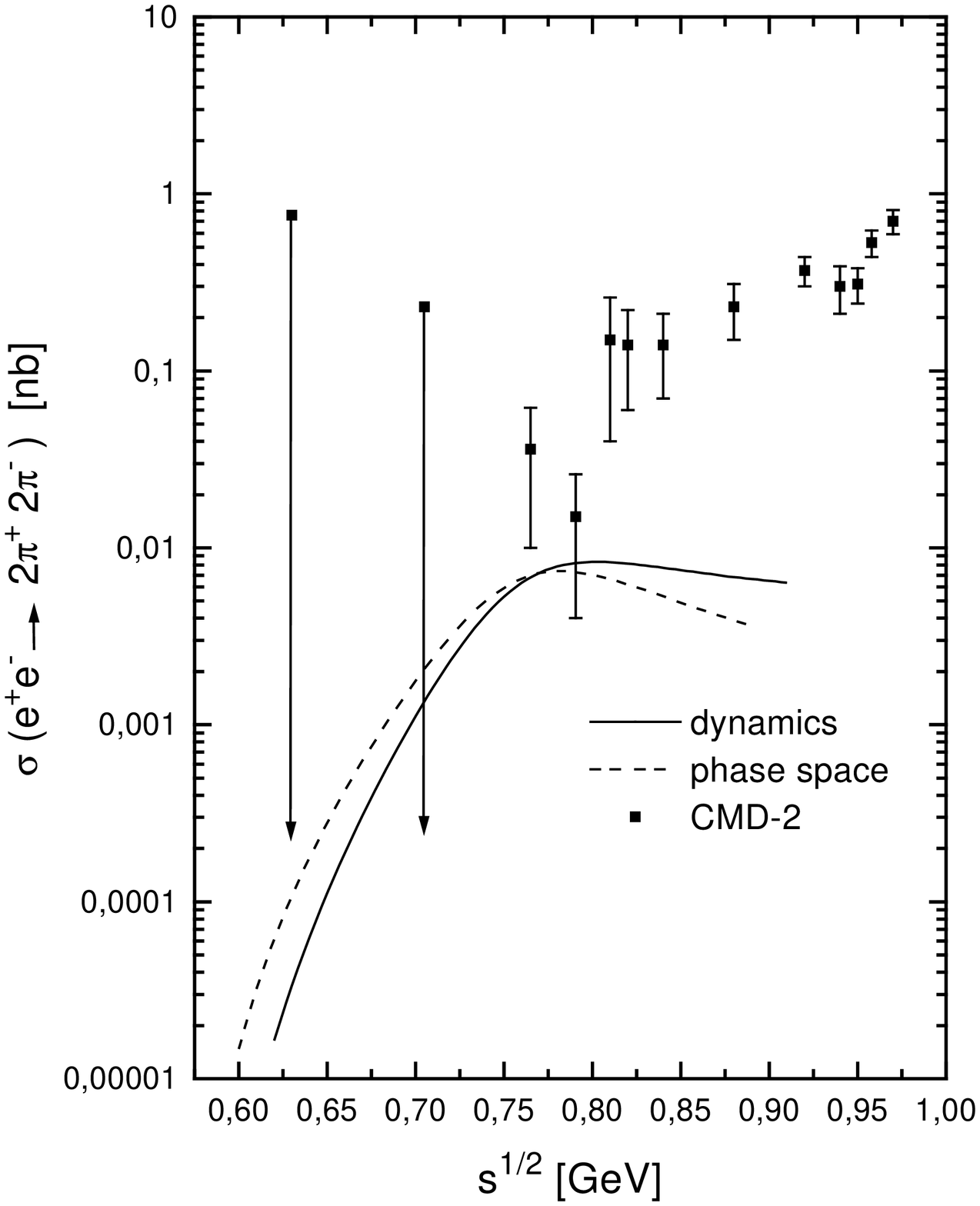}} \caption{The energy
dependence of the $e^+e^-\to\rho^0\to\pi^+\pi^-\pi^+\pi^-$
reaction cross section in the model based on the chiral Lagrangian
due to  Weinberg. Experimental points are from
Ref.~\protect\cite{cmd2}. \label{fig2}}
\end{figure}

\begin{figure}
\centerline {\epsfysize=7in \epsfbox{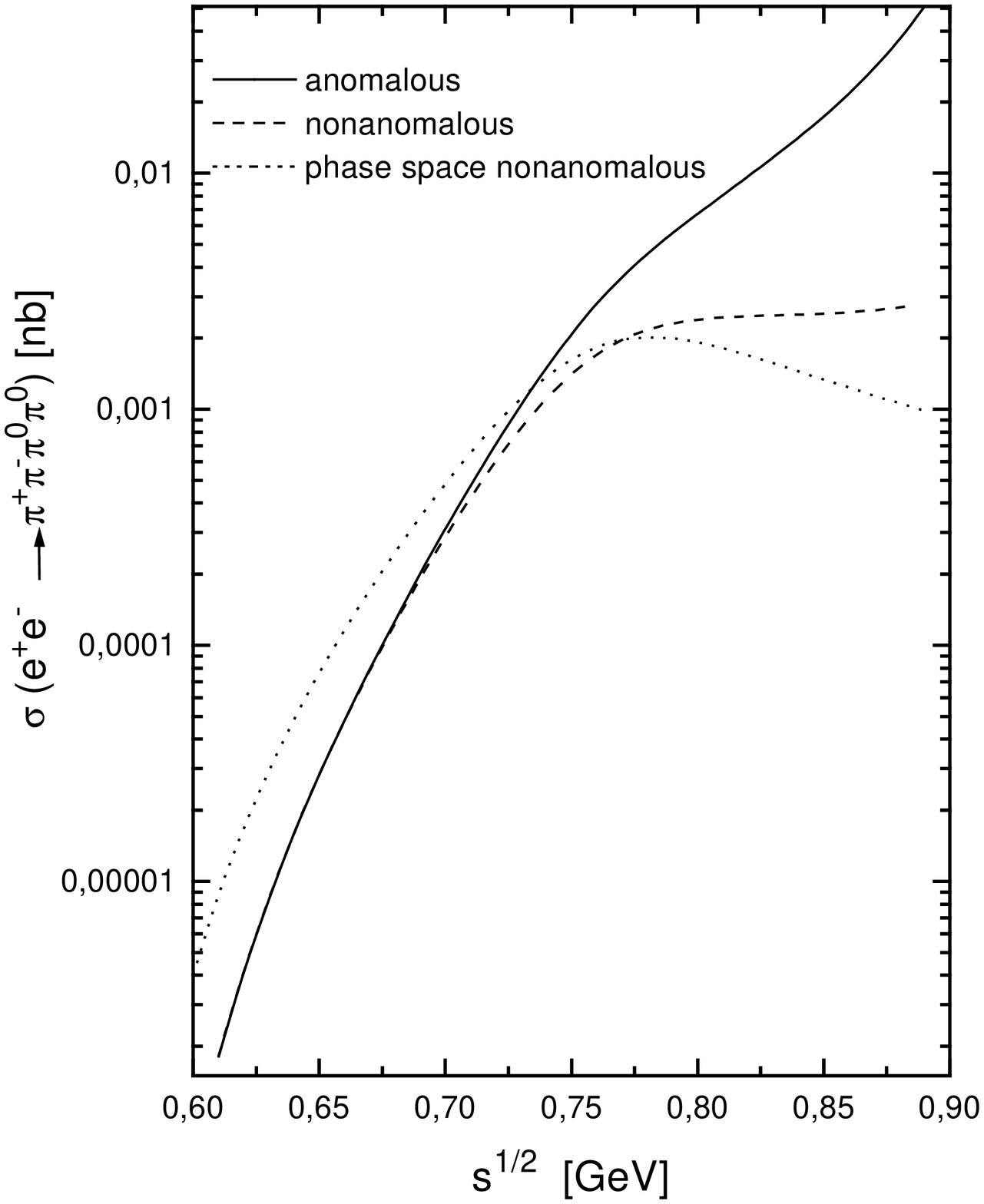}} \caption{The energy
dependence of the $e^+e^-\to\rho^0\to\pi^+\pi^-\pi^0\pi^0$
reaction cross section in the model based on the chiral Lagrangian
due to  Weinberg, added with the terms induced by the anomalous
Lagrangian of Wess and Zumino. \label{fig3}}
\end{figure}

The results of the evaluation of the $4\pi$ state production cross
section in the reaction $e^+e^-\to\rho^0\to4\pi$
\begin{equation}
\sigma_{e^+e^-\to\rho\to4\pi}(s)={12\pi m^3_\rho\Gamma_{\rho e^+e^-}(m_\rho)
\Gamma_{\rho\to4\pi}(E)\over E^3\left|D_\rho(s)\right|^2},
\label{rhoee}
\end{equation}
where $s=E^2$ is the square of the total center-of mass energy,
and $D_\rho(s)$ is obtained from Eq.~(\ref{proprho}) upon the
substitution $q^2\to s$, are plotted in Figs.~\ref{fig2} and
\ref{fig3}. Note that the values of the vector meson parameters
taken from Ref.~\cite{pdg} are used hereafter. The following
notations are such that
\begin{equation}
q(m_a,m_b,m_c)={1\over2m_a}\lambda^{1/2}(m^2_a,m^2_b,m^2_c),
\label{moment}
\end{equation}
with the function $\lambda$ given by the equation
\begin{equation}
\lambda(x,y,z)=x^2+y^2+x^2-2(xy+xz+yz),
\label{lambda}
\end{equation}
is the momentum of final particle $b$ (or $c$)
in the rest frame system of decaying particle $a$.

To demonstrate the effects of chiral dynamics,
also shown is the energy dependence of the cross section evaluated in the
model of pure phase space for the four pion decay. In this model, the
$4\pi$
partial width normalized to the width at the $\rho$ mass calculated in the
dynamical model, is given by the expression
\begin{equation}
\Gamma^{\rm LIPS}(\rho\to4\pi,s)=\Gamma(\rho\to4\pi,m^2_\rho){W_{4\pi}(s)
\over W_{4\pi}(m^2_\rho)},
\label{glips}
\end{equation}
where the four pion phase space volume is \cite{kumar,byck}
\begin{eqnarray}
W_{4\pi}(s)&=&{\pi^3\over16(2\pi)^8s^{3/2}N_s}\int_{(3m_\pi)^2}^{(\sqrt{s}
-m_\pi)^2}{ds_1\over s_1}\lambda^{1/2}(s,s_1,m^2_\pi) \nonumber\\
&&\times \int_{(2m_\pi)^2}^{(\sqrt{s_1} -m_\pi)^2}{ds_2\over
s_2}\lambda^{1/2}(s_1,s_2,m^2_\pi) \nonumber\\
&&\times\lambda^{1/2}(s_2,m^2_\pi,m^2_\pi). \label{w4pi}
\end{eqnarray}
In the above formula,  $N_s=4\mbox{ (2)}$ is the factor that takes
into account the identity of  final pions in the final state
$2\pi^+2\pi^-$($\pi^+\pi^-2\pi^0$), respectively. As the
evaluation shows,  the ratio $$R(s)=
\Gamma(\rho\to2\pi^+2\pi^-,s)/\Gamma^{\rm
LIPS}(\rho\to2\pi^+2\pi^-,s)$$ changes from 0.4 at $\sqrt{s}=650$
MeV to 1 at $\sqrt{s}=m_\rho$. As can be observed from the
figures, the rise of the $\rho\to4\pi$ partial width with the
energy increase is such fast that it compensates completely the
falling of the $\rho$ meson propagator and electronic width. Also
noticeable is the dynamical effect of the $\omega\pi^0$ production
threshold  in the decay $\rho^0\to\pi^+\pi^-2\pi^0$ at
$\sqrt{s}>850$ MeV which results from the anomaly induced
Lagrangian, see Fig.~\ref{fig3}. To quantify the abovementioned
effect of vanishing contribution of higher derivatives at the left
shoulder of the $\rho$ resonance it should be noted that the
difference between magnitude of
$\Gamma(\rho\to\pi^+\pi^-2\pi^0,s)$ with and without the term
originating from the anomaly induced Lagrangian, equal to $100\%$
at $\sqrt{s}=m_\rho$, diminishes rapidly with the energy decrease
amounting to $8\%$ at  $\sqrt{s}=700$ MeV and ¨ $0.3\%$ at
$\sqrt{s}=650$ MeV.
It should be pointed out that the evaluation of the partial widths
with the  nonrelativistic expressions for the $\rho\to4\pi$ amplitudes,
Eq.~(\ref{nonrel}), gives the values which deviate from those obtained
with the exact expressions, by the quantity ranging from 7 to 15 $\%$,
depending on the energy in the interval from 610 to 770 MeV.

As  is seen from Fig.~\ref{fig2}, the predictions of chiral
symmetry for the $e^+e^-\to2\pi^+2\pi^-$ reaction cross section do
not contradict to the three lowest experimental points of CMD-2
detector \cite{cmd2}. However, at $\sqrt{s}>800$ MeV one can
observe a substantial deviation between the predictions of the
Lagrangian (\ref{lwein}) and  the data of CMD-2. In all
appearance, this is due to the contribution of higher derivatives
and chiral loops neglected in the present work. It is expected
that the left shoulder of the $\rho$ is practically free of such
contributions, and by this reason it is preferable for studying
the dynamical effects of chiral symmetry. Note that even at
$\sqrt{s}=650$ MeV, where the contribution of higher derivatives
is negligible, one can hope to gather one event of the reaction
$e^+e^-\to2\pi^+2\pi^-$ per day,  and up to 10 events of this
reaction per day at $\sqrt{s}=700$ MeV, provided the luminosity
$L=10^{32}\mbox{cm}^{-2}\mbox{s}^{-1}$ is achieved, i.e. to have a
factory for a comprehensive study of the chiral dynamics of
many-pion systems.

Due to helicity conservation, $\rho$ meson is produced in the
states with the spin projections $\lambda=\pm1$ on the $e^+e^-$
beam axes characterized by the unit vector ${\bf n}_0$. The latter
is assumed to be directed along the z axes. Then, using the
expressions for the total $\rho\to4\pi$ amplitudes, one can obtain
the angular distributions for the final pions. The are expected to
be cumbersome. However, the good approximation for these
distributions are obtained from the approximate nonrelativistic
expression Eq.~(\ref{nonrel}). The specific formulas are collected
in the Appendix.

\subsection{$\rho\to4\pi$ in $\tau$ decays}
\label{susectau}

\begin{figure}
\centerline {\epsfysize=7in \epsfbox{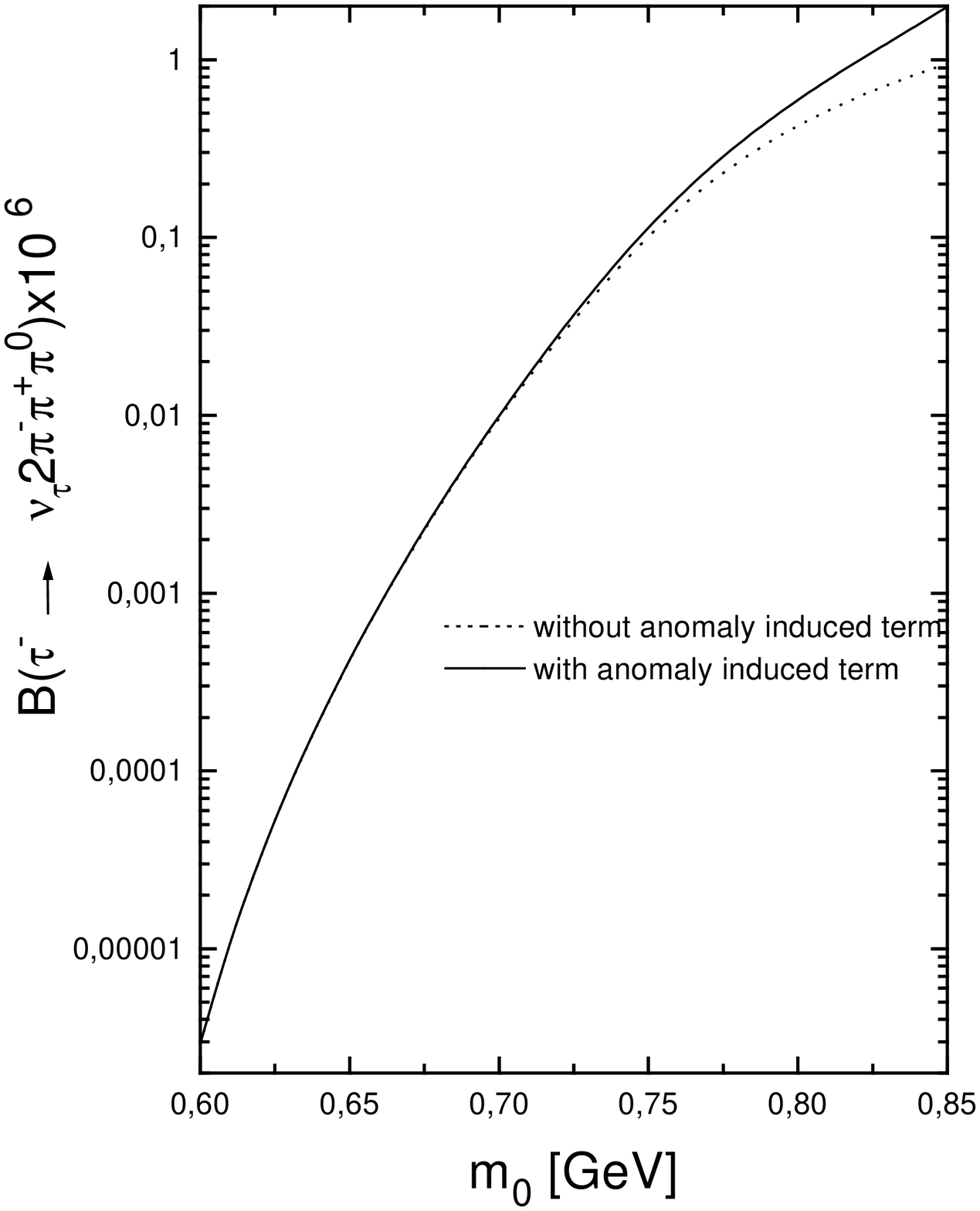}}
\caption{The dependence of the branching ratio of the decay
$\tau^-\to\nu_\tau 2\pi^-\pi^+\pi^0$ on the invariant mass of the four
pion system, see Eq.~(\protect\ref{widtau}).
\label{fig4}}
\end{figure}

\begin{figure}
\centerline {\epsfysize=7in \epsfbox{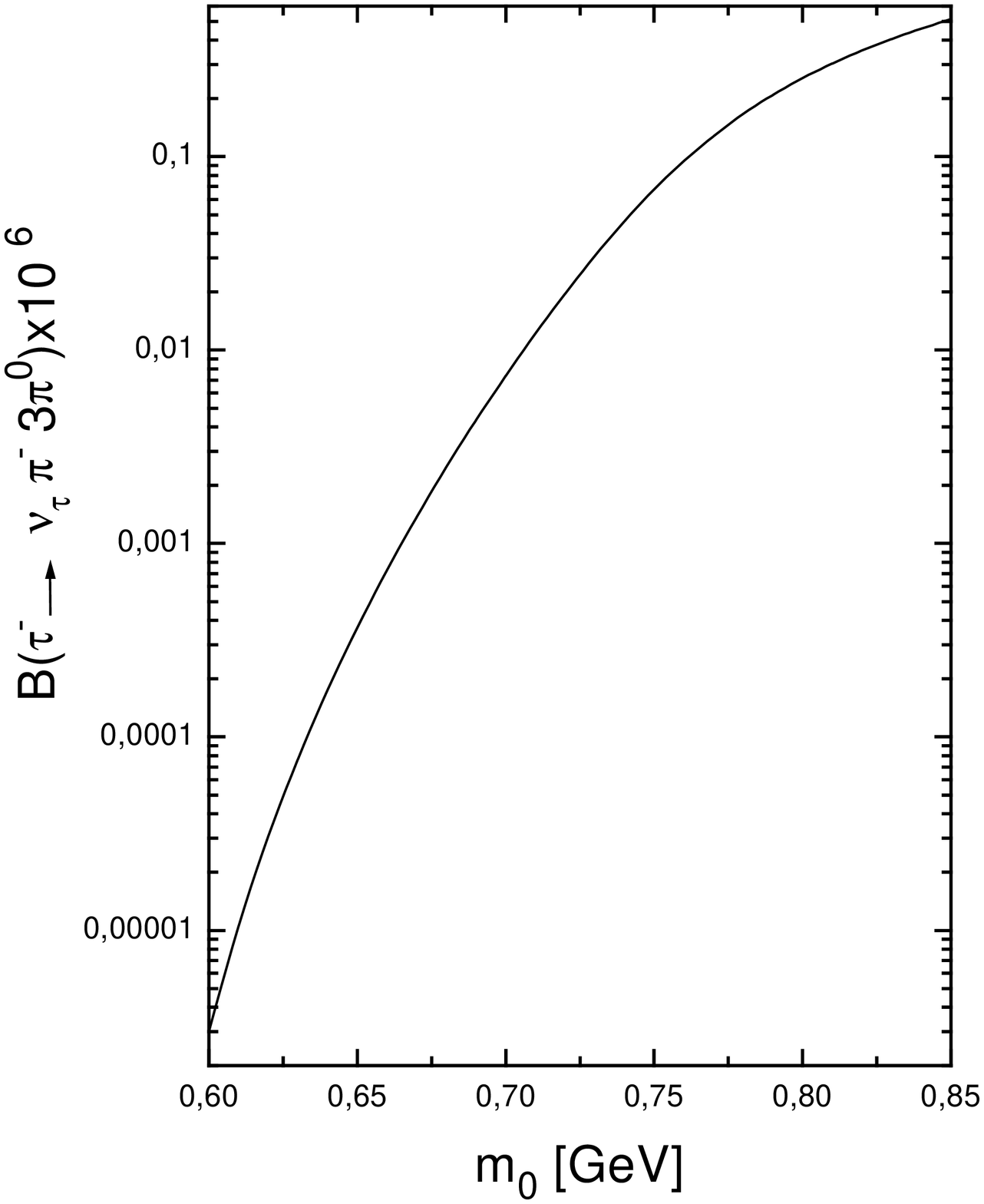}}
\caption{The same as in Fig.~\protect\ref{fig4}, but for the decay
$\tau^-\to\nu_\tau\pi^-3\pi^0$.
\label{fig5}}
\end{figure}

Based on the vector current conservation, the partial width of the decay
$\tau^-\to\nu_\tau(4\pi)^-$ \cite{tsai,gilman}
can be written as the integral over the
invariant mass of the four pion state $m$, extended up to some mass $m_0$,
whose maximal value is $m_{0\rm max}=m_\tau$:
\begin{eqnarray}
B_{\tau^-\to\nu_\tau(4\pi)^-}(m_0)&=&T_\tau\int_{4m_\pi}^{m_0}dm
\frac{2m^2\Gamma_{\tau^-\to\nu_\tau\rho^-}(m)}{\pi
|D_\rho(m^2)|^2}     \nonumber\\
&&\times\Gamma_{\rho^-\to(4\pi)^-}(m),
\label{widtau}
\end{eqnarray}
where $T_\tau$ and
\begin{eqnarray}
\Gamma_{\tau^-\to\nu_\tau\rho^-}(m)&=&{G^2_F\cos^2\theta_C\over
8\pi f^2_\rho}m^3_\tau m^2_\rho
\left(1-{m^2\over m^2_\tau}\right)^2  \nonumber\\
&&\times\left(1+2{m^2\over m^2_\tau}\right)
\label{tauronu}
\end{eqnarray}
are, respectively, the lifetime of $\tau$ lepton and its partial
width of the decay $\tau^-\to\nu_\tau\rho^-$ \cite{tsai}, with $m$
being the invariant mass of the four pion state. Using the
numerical values of the $\rho\to4\pi$ decay widths, one can
evaluate the branching ratios of the four pion $\tau$ decays for
various values of the upper invariant mass $m_0$ of the latter.
The results of the evaluation of the branching ratios of the
decays $\tau^-\to\nu_\tau2\pi^-\pi^+\pi^0$ and
$\tau^-\to\nu_\tau\pi^-3\pi^0$ for the values of the invariant
mass of the four pion system from 600 to 850 MeV are plotted in
Figs.~\ref{fig4} and \ref{fig5}, respectively. In particular,
taking $m_0=740$ MeV one obtains \widetext
\begin{equation}
B(\tau^-\to\nu_\tau2\pi^-\pi^+\pi^0,740\mbox{MeV})=
\left\{7.6\times10^{-8}(\mbox{ without anomaly induced term})\atop
8.4\times10^{-8}(\mbox{ with anomaly induced term})\right.,
\label{tau7ch}
\end{equation}
\narrowtext
and
\begin{equation}
B(\tau^-\to\nu_\tau3\pi^0\pi^-,740\mbox{MeV})=
4.6\times10^{-8}.
\label{tau7ne}
\end{equation}
Corresponding values for the upper integration mass $m_0=640$ MeV
are
\widetext
\begin{equation}
B(\tau^-\to\nu_\tau2\pi^-\pi^+\pi^0,640\mbox{MeV})=
\left\{2.895\times10^{-10}(\mbox{ without anomaly induced term})\atop
2.900\times10^{-10}(\mbox{ with anomaly induced term})\right.,
\label{tau6ch}
\end{equation}
\narrowtext
and
\begin{equation}
B(\tau^-\to\nu_\tau3\pi^0\pi^-,640\mbox{MeV})=
1.8\times10^{-10}.
\label{tau6ne}
\end{equation}
The comparison of Eqs.~(\ref{tau7ch}) and (\ref{tau6ch}), and both
curves in Fig.~\ref{fig4} again demonstrates that the
contributions of higher derivatives represented by the terms
induced by the anomalous Lagrangian of Wess and Zumino vanish
rapidly with the decreasing of mass. Unfortunately, the domains in
the low four pion invariant mass where the effects of chiral
dynamics are clean, are hardly accessible with $\tau$ factories.
Indeed, guided by the expression for the cross section of the
$\tau$ lepton pair production in $e^+e^-$ annihilation
\begin{equation}
\sigma_{\tau^+\tau^-}(s)={4\pi\alpha^2\over3s}
\sqrt{1-{4m^2_\tau\over s}}\left(1+2{m^2_\tau\over s}\right),
\label{ee2tau}
\end{equation}
one can find that up to $N=25\times10^7$ $\tau$ lepton pairs with the total
energy $\sqrt{s}=m_{\psi(2S)}$  can be produced per season
at $\tau$-charm factory with expected luminosity
$L=10^{34}$ cm$^{-2}$ s$^{-1}$ \cite{pdg}.
This implies that one can detect only
from 2 to 4 events per season in the four pion mass range below 700 MeV.
Nevertheless, the event counting rate  rises rapidly with the increase of
the upper integration mass $m_0$ in Eq.~(\ref{widtau}), reaching, at
$m_0=m_\rho$, the figure about  60 to 120 events per season, depending on
the charge combination of the final pions.

\subsection{The decay $\rho\to4\pi$ in photoproduction, $\pi N\to\rho\pi N$,
and so on.}
\label{susecph}

\begin{figure}
\centerline {\epsfysize=7in \epsfbox{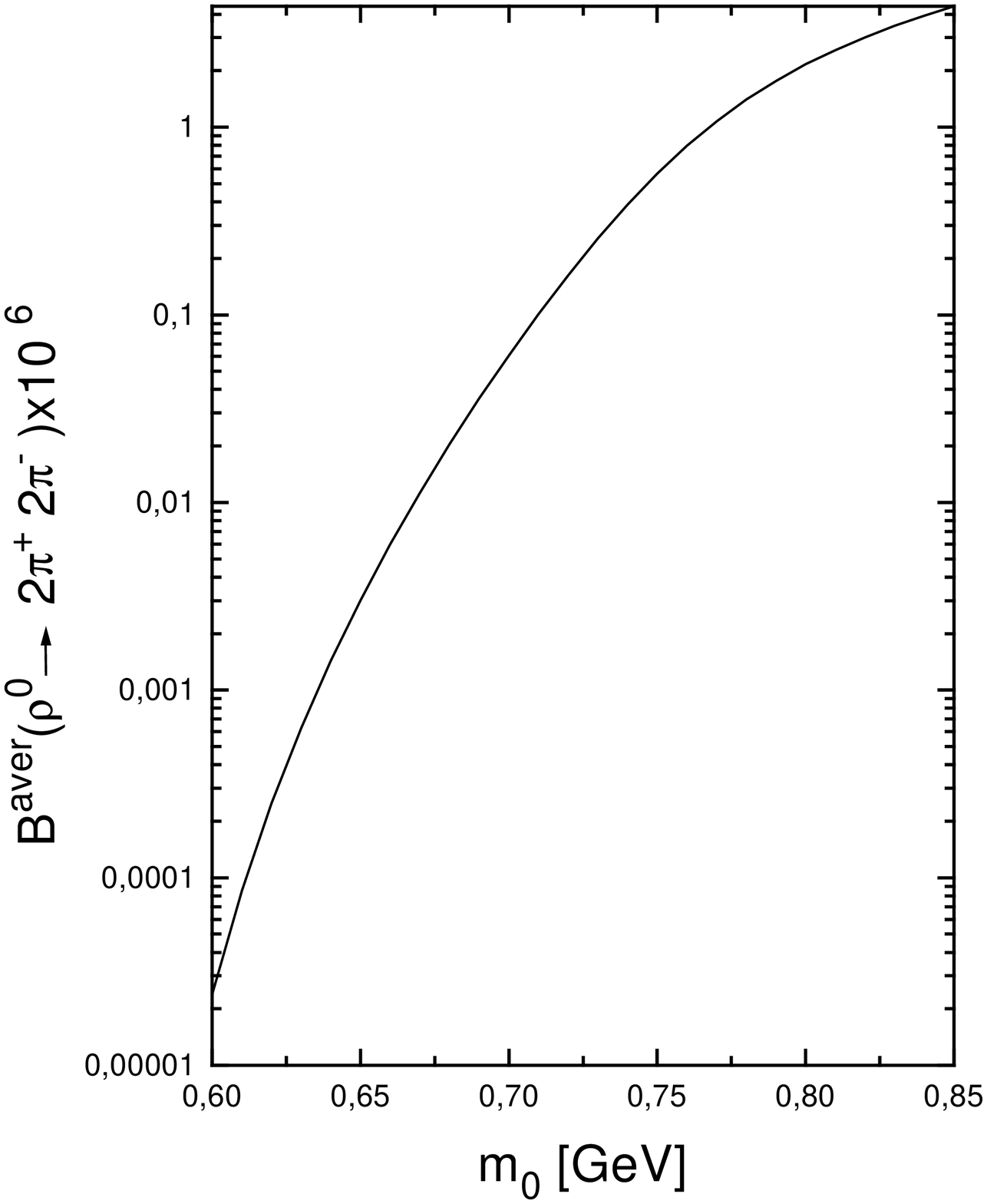}}
\caption{The dependence of the branching ratio of the decay
$\rho^0\to2\pi^+2\pi^-$ averaged over the invariant mass of the four
pion system, see Eq.~(\protect\ref{bm}).
\label{fig6}}
\end{figure}

\begin{figure}
\centerline {\epsfysize=7in \epsfbox{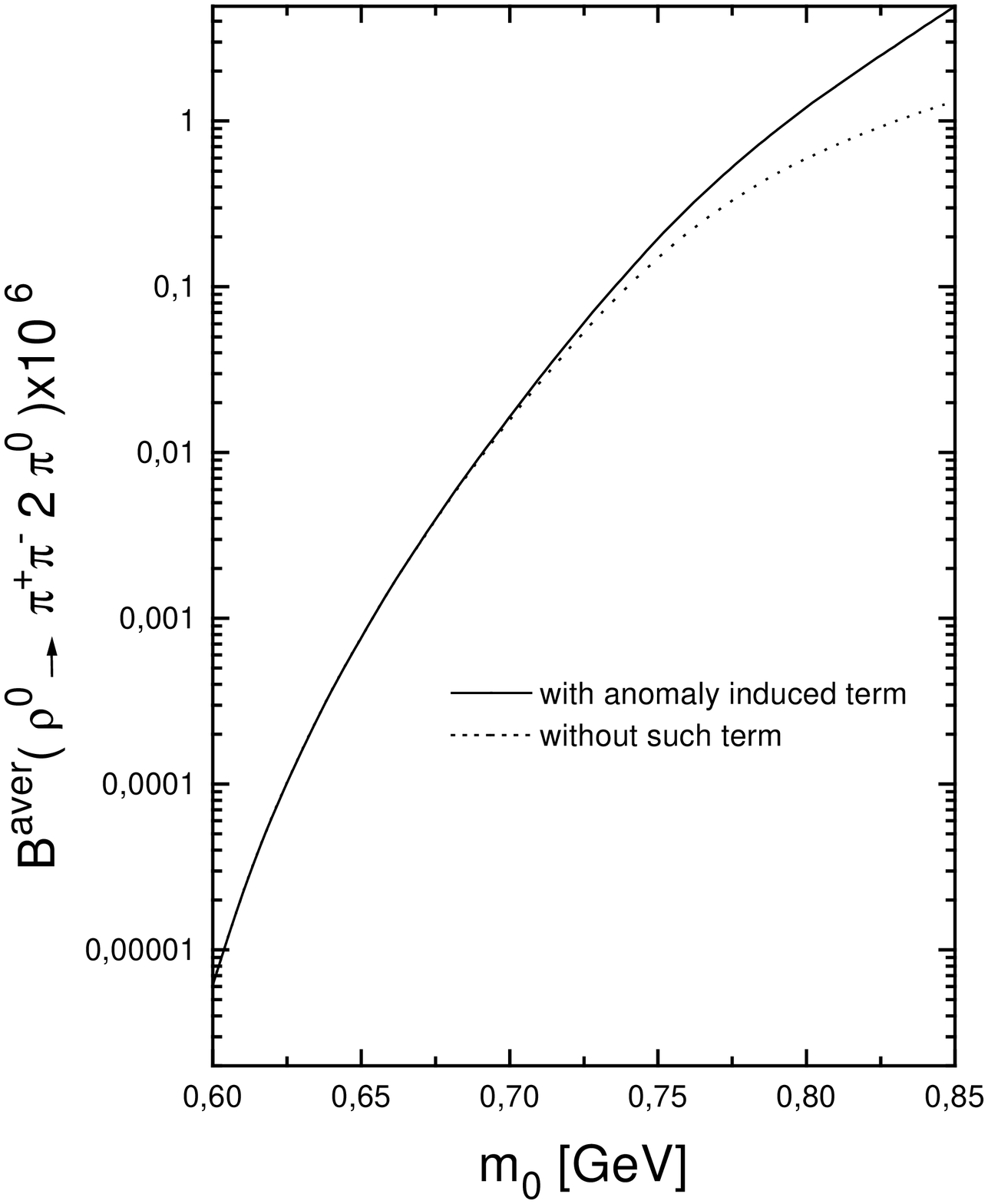}}
\caption{The same as in Fig.~\protect\ref{fig6} but for the decay
$\rho^0\to\pi^+\pi^-2\pi^0$.
\label{fig7}}
\end{figure}

To characterize the possibility of the study of the $\rho\to4\pi$ decays
in photoproduction, we calculate the quantity
\begin{equation}
B^{\rm aver}_{\rho\to4\pi}(m_0)=
{2\over\pi}\int_{4m_\pi}^{m_0}dm{m^2\Gamma_{\rho\to4\pi}(m)\over
|D_\rho(m^2)|^2},
\label{bm}
\end{equation}
which is the average of the branching ratio over the invariant mass of the
four pion state. In the limit $m_0\to\infty$, Eq.~(\ref{bm})
serves as the definition of the branching
ratio in case of a wide resonance.
Equation~(\ref{bm}) should be confronted with
the familiar definition of the branching ratio at the $\rho$ mass
\begin{equation}
B_{\rho\to4\pi}(m_\rho)=\Gamma_{\rho\to4\pi}/\Gamma_\rho,
\label{bmm}
\end{equation}
which results from Eq.~(\ref{bm}) upon the replacement
$m\Gamma_\rho/\pi|D_\rho(m^2)|^2\to\delta(m^2-m^2_\rho)$ valid in the limit
of narrow width. With the partial widths evaluated here one finds
\begin{equation}
B_{\rho^0\to2\pi^+2\pi^-}(m_\rho)=5.9\times10^{-6}, \label{cha0}
\end{equation}
and
\widetext
\begin{equation}
B_{\rho^0\to\pi^+\pi^-2\pi^0}(m_\rho)=
\left\{1.6\times10^{-6}(\mbox{ without anomaly induced term})\atop
2.9\times10^{-6}(\mbox{ with anomaly induced term})\right..
\label{neu0}
\end{equation}
\narrowtext The results of plotting the quantity $B^{\rm
aver}_{\rho\to4\pi}(m_0)$ are shown in Fig.~\ref{fig6} and
\ref{fig7}. In particular, the evaluation gives $B^{\rm
aver}_{\rho^0\to2\pi^+2\pi^-}(m_0)=4.4\times10^{-6} \mbox{,
}6.1\times10^{-8}$, and $1.4\times10^{-9}$ at $m_0=850$, 700, and
640 MeV, respectively. In the case of other four pion decay mode
of the $\rho^0$ the results are the following. In the model with
the vanishing term induced by the anomalous Lagrangian of Wess and
Zumino one obtains $B^{\rm
aver}_{\rho^0\to\pi^+\pi^-2\pi^0}(m_0)=1.3\times10^{-6}\mbox{,
}1.58 \times10^{-8}$, and $3.66\times10^{-10}$ at $m_0=850$, 700,
and 640 MeV. In the model that includes the above term, one
obtains $B^{\rm
aver}_{\rho^0\to\pi^+\pi^-2\pi^0}(m_0)=4.9\times10^{-6}\mbox{, }
1.65\times10^{-8}$, and $3.63\times10^{-10}$ at the same
respective values of $m_0$. As is expected, the branching ratios
in the two mentioned models converge to each other in view of the
rapid vanishing of the contributions due to the terms with higher
derivatives. The difference between the two definitions of the
branching ratio is seen upon comparison of $B^{\rm
aver}_{\rho\to4\pi}(m_0=850\mbox{ MeV})$ evaluated for various
charge combinations of the final pions, with Eqs.~(\ref{cha0}) and
(\ref{neu0}).

With the total number of $\rho$ mesons $N\simeq6\times10^{9}$
expected to be produced on nucleon at the Jefferson Laboratory
"photon factory" \cite{dzierba} one may hope to observe about 100, 360 events
of the $\rho$ decays into the states $\pi^+\pi^-2\pi^0$, $2\pi^+2\pi^-$,
respectively, in the mass range $m_0<700$ MeV where the effects of chiral
dynamics are most clean. The photoproduction on heavy nuclei results in
increasing the number of produced $\rho$ mesons faster than $A^{2/3}$,
where $A$ is the atomic weight. A generally adopted behavior is in accord with
the behavior $A^{0.8-0.95}$ \cite{leith}.
Thus  the photoproduction of the four
pion states on heavy nuclei would give the possibility of the high statistics
study of the effects of chiral dynamics in the four pion decays of the
$\rho(770)$.
It should be recalled once more that the counting
rate rises rapidly with the increase of $m_0$.

The conclusions about the angular distributions of the final pions
with zero net charge in photoproduction are the following. Of
course, their general expression should be deduced from the full
decay amplitudes which can be found in Sec.~\ref{sec2}, together
with the detailed form of the photoproduction mechanism. The
qualitative picture, however, can be obtained upon assuming
$s$-channel helicity conservation to be a good selection rule for
the photoproduction reactions \cite{leith,bauer}. Then in the
helicity reference frame characterized as the frame where the
$\rho$ is at rest, while its spin quantization axes is directed
along the $\rho$ momentum in the center-of-mass system, the
expressions for the angular distributions coincide with the
corresponding expressions for the production of these states in
$e^+e^-$ annihilation that can be found in the Appendix.  Since, at
high energies, the direction of the final $\rho$ momentum lies at
the scattering angle less than $0.5^\circ$ in the case of the
photoproduction on heavy nuclei, the vector ${\bf n}_0$ can be
treated as pointed along the photon beam direction.

\begin{figure}
\centerline {\epsfysize=7in \epsfbox{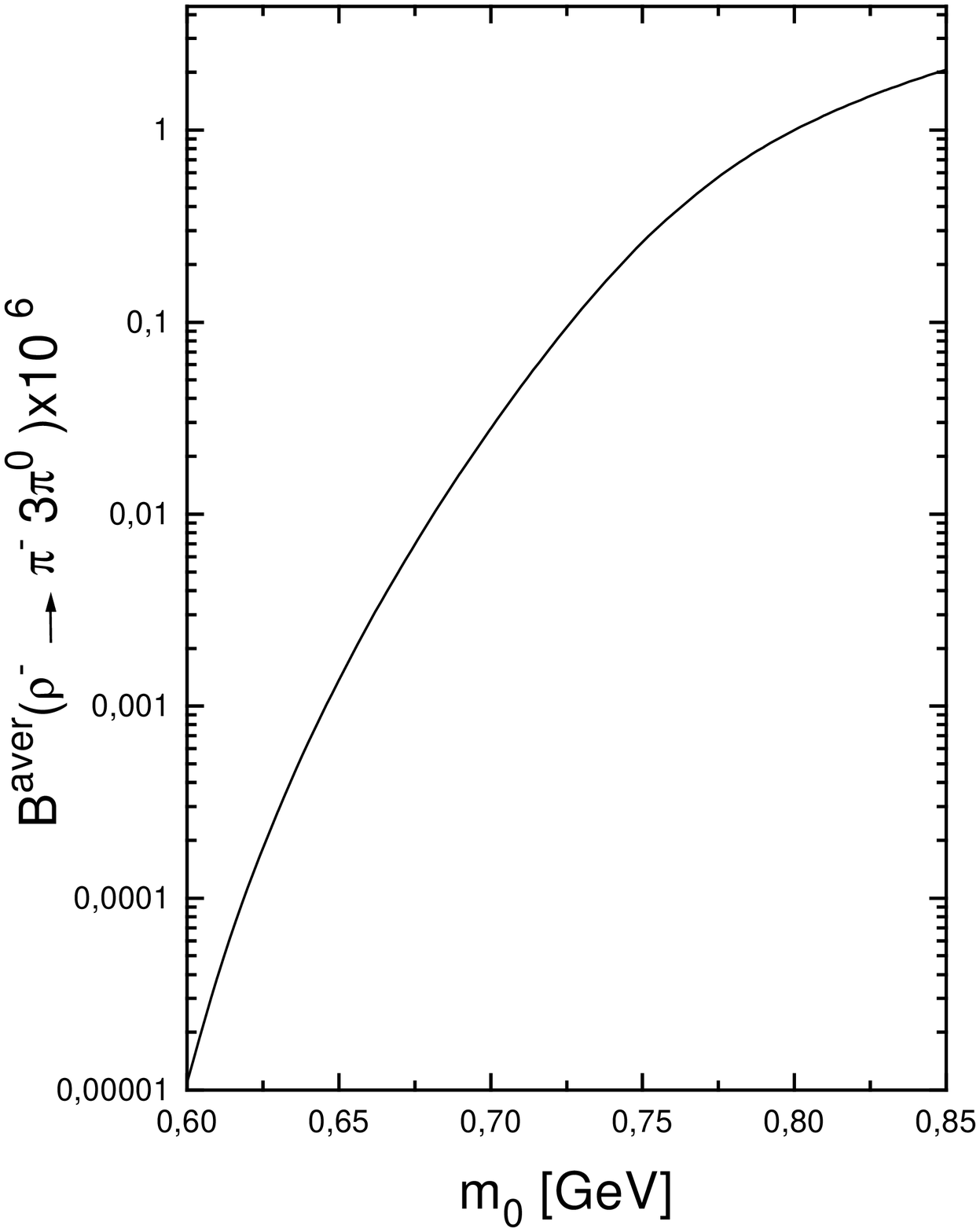}} \caption{The same
as in Fig.~\protect\ref{fig6} but for the decay
$\rho^-\to\pi^-3\pi^0$. \label{fig8}}
\end{figure}

\begin{figure}
\centerline {\epsfysize=7in \epsfbox{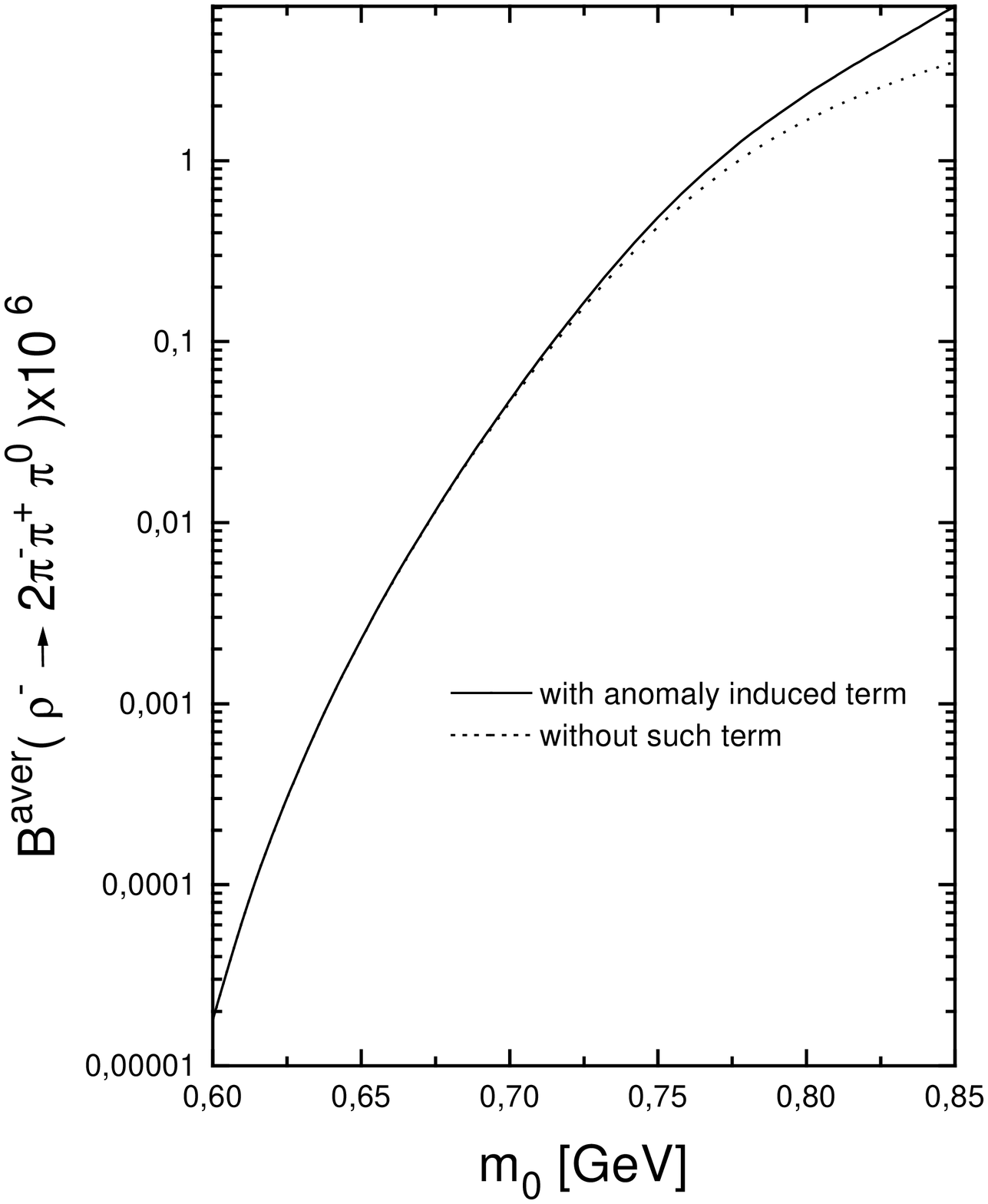}} \caption{The same
as in Fig.~\protect\ref{fig6} but for the decay
$\rho^-\to2\pi^-\pi^+\pi^0$. \label{fig9}}
\end{figure}

Note that another peripheral reactions can provide a sufficiently
intense source of $\rho$ mesons. For example, the diffractive
production of the $\rho\pi$ state in $\pi N$ collisions are
currently under study with the VES detector in Protvino. The
regions of the four pion invariant mass spectrum larger than
$m_\rho$, namely, $m_0\simeq$ 850 MeV where $B^{\rm
aver}(\rho\to4\pi,m_0)\sim 10^{-5}$ should be included to measure
the $\rho\to4\pi$ branching ratio reliably. As is explained in
Introduction, this would require the inclusion of the
contributions of $a_1$ meson and higher derivatives to the total
amplitude. Nevertheless, the results of the present paper shown in
Figs.~\ref{fig8} and \ref{fig9}, obtained upon neglecting the
latter contributions can be regarded as a guess in the
experimental work in this direction.

\section{The decay $\omega\to5\pi$}
\label{sec4}

\begin{figure}
\centerline {\epsfysize=7in \epsfbox{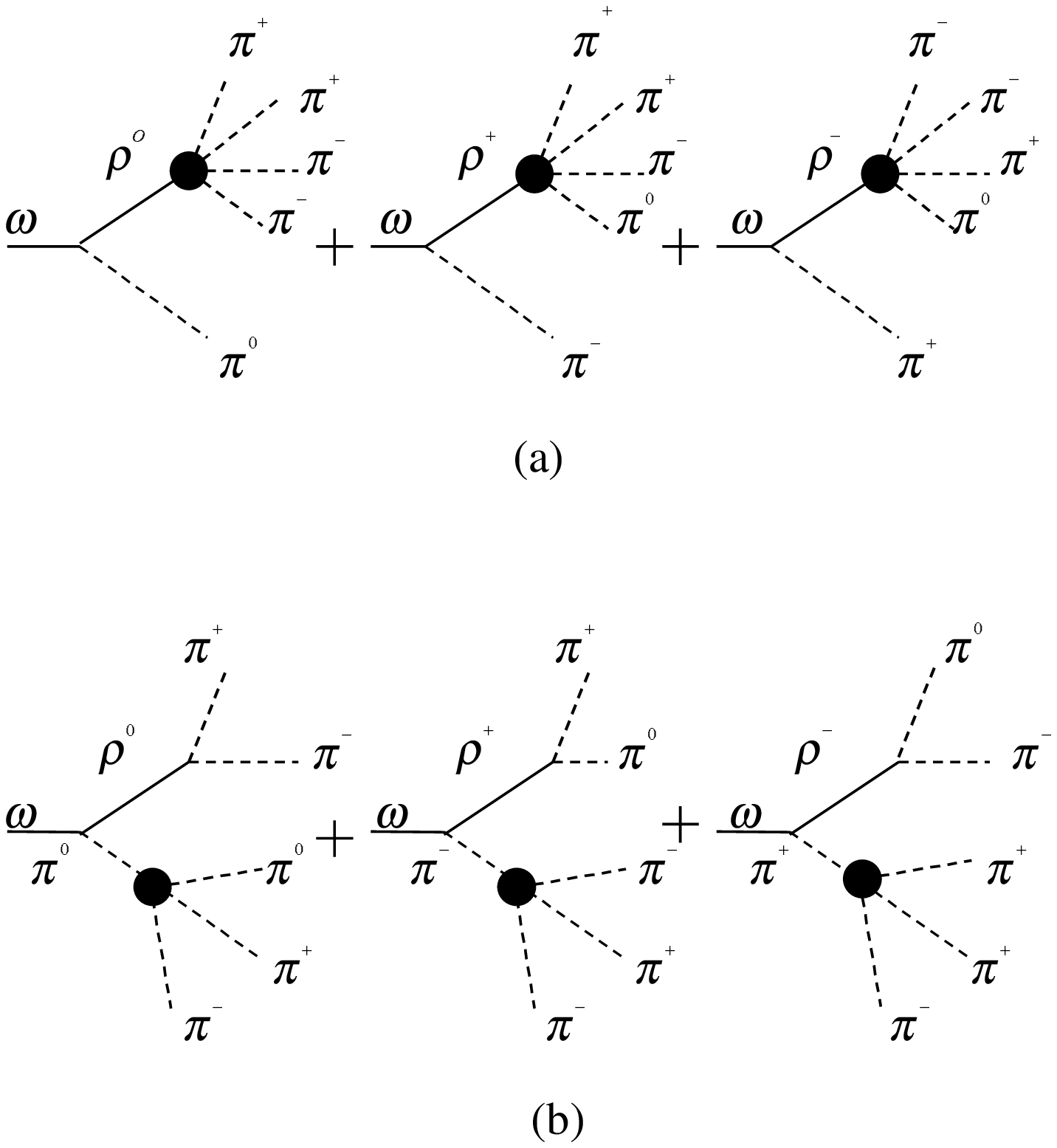}} \caption{The
diagrams describing the amplitudes of the decays
$\omega\to\pi^+\pi^-\pi^+\pi^-\pi^0$. The shaded circles in the
set (a) denote the whole set of the $\rho\to4\pi$ diagrams shown
in Figs.~\protect\ref{fig1}(b), (c). The shaded circles in the
$\pi\to3\pi$ vertices in the set (b) refer to the sum of diagrams
shown in Fig.~\protect\ref{fig1}(a). The symmetrization over
momenta of identical pions emitted from different vertices is
meant. The diagrams for the decay
$\omega\to\pi^+\pi^-\pi^0\pi^0\pi^0$ are obtained from those shown
upon the evident replacements. \label{fig10}}
\end{figure}

One may convince oneself that the $\omega\to\rho\pi\to5\pi$ decay
amplitude unambiguously results from the anomaly induced
Lagrangian (\ref{wz}). This amplitude is represented by the
diagrams shown in Fig.~\ref{fig10}. As one can foresee, its
general expression looks cumbersome. However, it can be
considerably simplified upon noting  that due to the low pion
momentum,  $|{\bf q}_\pi|\simeq0.5m_\pi$, the nonrelativistic
expressions Eq.~(\ref{nonrel}) for the $\rho\to4\pi$ decay
amplitudes in the diagrams Fig.~\ref{fig10}(a) are valid with the
accuracy $5\%$ in the $4\pi$ mass range relevant for the present
purpose \cite{ach99}. This accuracy is estimated from the direct
evaluation of the $\rho\to4\pi$ branching ratios with the exact
$\rho\to4\pi$ decay formulas given in Sec.~\ref{sec2}, and with
the approximate ones, Eq.~(\ref{nonrel}). The evaluation shows
that the results differ by approximately 10 $\%$ in the width.
Likewise, the expression for the combination $D^{-1}_\pi
M(\pi\to3\pi)$ standing in the expression for the diagrams in
Fig.~\ref{fig10}(b) can be replaced, with the same accuracy, by
$-(8m^2_\pi)^{-1}$ times the nonrelativistic $\pi\to3\pi$
amplitudes in Eq.~(\ref{4pinr}). First, using Eq.~(\ref{nonrel})
one obtains the expression for the sum of the diagrams shown in
Fig.~\ref{fig10}(a). Second, using Eq.~(\ref{4pinr}), one obtains
the expression for the sum of the diagrams shown in
Fig.~\ref{fig10}(b). Note that the contribution of the diagrams
Fig.~\ref{fig10}(b) was neglected  in Ref.~\cite{ach99}. The final
expressions for the $\omega\to5\pi$ decay amplitudes, upon
neglecting the terms of the order of $O(|{\bf q}^4_\pi/m^4_\pi)$
or higher,  can be represented in the form:
\widetext
\begin{eqnarray}
M(\omega\to2\pi^+2\pi^-\pi^0)&=&{N_cg_{\rho\pi\pi}g^2\over8(2\pi)^2
f^3_\pi}\varepsilon_{\mu\nu\lambda\sigma}q_\mu\epsilon_\nu
\left\{(1+P_{12})q_{1\lambda}\left[{(q_2+3q_4)_\sigma\over
D_\rho(q-q_1)} -{2q_{4\sigma}\over D_\rho(q_1+q_4)}\right]\right.
\nonumber\\
&&\left.-(1+P_{35})q_{3\lambda}\left[{(q_5+3q_4)_\sigma\over
D_\rho(q-q_3)} -{2q_{4\sigma}\over D_\rho(q_3+q_4)}\right]\right.
\nonumber\\ &&\left.-
(1+P_{12})(1+P_{35})q_{3\lambda}\left[{2q_{4_\sigma}\over
D_\rho(q-q_4)} +{q_{1\sigma}\over D_\rho(q_1+q_3)}\right]\right\},
\label{om1pi0}
\end{eqnarray}
\narrowtext
with the final momentum  assignment according to
$\pi^+(q_1)\pi^+(q_2)\pi^-(q_3)\pi^-(q_5)\pi^0(q_4)$, and
\widetext
\begin{eqnarray}
M(\omega\to\pi^+\pi^-3\pi^0)&=&{N_cg_{\rho\pi\pi}g^2\over8(2\pi)^2
f^3_\pi}(1-P_{12})(1+P_{34}+P_{35})
\varepsilon_{\mu\nu\lambda\sigma}q_\mu\epsilon_\nu q_{1\lambda}
\nonumber\\ &&\times\left\{q_{3\sigma}\left[{1\over D_\rho(q-q_3)}
-{1\over D_\rho(q_1+q_3)}\right]\right.\nonumber\\
&&\left.-q_{2\sigma}\left[{4\over3D_\rho(q-q_1)} -{1\over
2D_\rho(q_1+q_2)}\right]\right\}, \label{om3pi0}
\end{eqnarray}
\narrowtext
with the final momentum  assignment according to
$\pi^+(q_1)\pi^-(q_2)\pi^0(q_3)\pi^0(q_4)\pi^0(q_5)$.
In both above formulas, $\epsilon_\nu$, $q_\mu$ stand for four-vectors
of polarization and momentum of $\omega$ meson. Note that the first term
in each square bracket refers to the specific diagram
shown in Fig.~\ref{fig10}(a) while the second one does to the diagram
shown in Fig.~\ref{fig10}(b).

Yet even in this simplified form the expressions for the
$\omega\to5\pi$ amplitudes are not easy to use for evaluation of
the branching ratios. To go further, one should  note the
following. One can check that the invariant mass of the $4\pi$
system on which the contribution of the diagrams shown in
Fig.~\ref{fig10}(a) depends, changes in the very narrow range
$558\mbox{ MeV}<m_{4\pi}<642\mbox{ MeV}$. Hence, one can set it in
all the $\rho$ propagators standing as the first terms in all
square brackets in Eqs.~(\ref{om3pi0}) and (\ref{om1pi0}), with
the accuracy $20\%$ in width, to the "equilibrium" value
${\overline{m^2_{4\pi}}}^{1/2}=620$ MeV evaluated for the
pion energy $E_\pi=m_\omega/5$ which gives the
dominant contribution. The same is true for the invariant mass of
the pion pairs on which the $\rho$ propagators standing as the
last terms in square brackets of the above expressions, depend.
This invariant mass varies in the narrow range $280\mbox{
MeV}<m_{2\pi}<360\mbox{ MeV}$. With the same accuracy, one can set
it to $\overline{m^2_{2\pi}}^{1/2}=295$ MeV in all relevant
propagators. On the other hand, the amplitude of the process
$\omega\to\rho^0\pi^0\to(2\pi^+2\pi^-)\pi^0$ is
\begin{eqnarray}
M[\omega\to\rho^0\pi^0\to(2\pi^+2\pi^-)\pi^0]
&=&4{N_cg_{\rho\pi\pi}g^2\over8(2\pi)^2f^3_\pi}
\nonumber\\
&&\times
\varepsilon_{\mu\nu\lambda\sigma}q_\mu\epsilon_\nu(q_1+q_2)_\lambda
\nonumber\\
&&\times{q_{4\sigma}\over D_\rho(q-q_4)},
\label{ampli}
\end{eqnarray}
where the momentum assignment is the same as in
Eq.~(\ref{om1pi0}). The other relevant amplitude corresponding to
the first diagram in Fig.~\ref{fig10}(b) is \widetext
\begin{eqnarray}
M[\omega\to\rho^0\pi^0\to(\pi^+\pi^-)(\pi^+\pi^-\pi^0)]&=&
{N_cg_{\rho\pi\pi}g^2\over8(2\pi)^2f^3_\pi}
\varepsilon_{\mu\nu\lambda\sigma}q_\mu\epsilon_\nu
\nonumber\\
&&\times(1+P_{12})(1+P_{35}){q_{1\lambda}q_{3\sigma}\over D_\rho(q_1+q_3)}
\label{ampli1}
\end{eqnarray}
\narrowtext Taking into account the above consideration concerning
the invariant masses, one can replace all the $\rho$ propagators
standing as the first terms in square parentheses of
Eq.~(\ref{om1pi0}) with $1/D_\rho(q-q_4)$ one. In the same manner,
all the $\rho$ propagators standing as the second terms in square
parentheses can be replaced with $1/D_\rho(q_1+q_3)$. Then the
comparison of Eqs.~(\ref{om1pi0}), (\ref{ampli}), and
(\ref{ampli1}) shows that
\begin{eqnarray}
M(\omega\to2\pi^+2\pi^-\pi^0)&\approx&{5\over2}
M[\omega\to\rho^0\pi^0\to(2\pi^+2\pi^-)\pi^0]   \nonumber\\
&&\times\left[1-{D_\rho(\overline{m^2_{4\pi}})
\over2D_\rho(\overline{m^2_{2\pi}})}\right], \label{ampli2}
\end{eqnarray}
where we replace the ratio $D_\rho(q-q_4)/D_\rho(q_1+q_3)$ with
the ratio $D_\rho(\overline{m^2_{4\pi}})/
D_\rho(\overline{m^2_{2\pi}})$ evaluated at the "equilibrium"
point.  The same treatment shows that
\begin{eqnarray}
M(\omega\to\pi^+\pi^-3\pi^0)&\approx&{5\over2}
M[\omega\to\rho^+\pi^-\to(\pi^+3\pi^0)\pi^-]   \nonumber\\
&&\times\left[1-{D_\rho(\overline{m^2_{4\pi}})
\over2D_\rho(\overline{m^2_{2\pi}})}\right],
\label{ampli3}
\end{eqnarray}
where
\begin{eqnarray}
M[\omega\to\rho^+\pi^-\to(\pi^+3\pi^0)\pi^-]
&=&-4{N_cg_{\rho\pi\pi}g^2\over8(2\pi)^2f^3_\pi}
\nonumber\\
&&\times
{\varepsilon_{\mu\nu\lambda\sigma}q_\mu\epsilon_\nu q_{1\lambda}q_{2\sigma}
\over D_\rho(q-q_2)},
\label{ampli4}
\end{eqnarray}
and the final momenta assignment is the same as in
Eq.~(\ref{om3pi0}). The numerical values of
$\overline{m^2_{4\pi}}^{1/2}$ and $\overline{m^2_{2\pi}}^{1/2}$
found above are such that  the correction factor in parentheses of
Eqs.~(\ref{ampli2}) and (\ref{ampli3}) amounts to $20\%$ in
magnitude. In what follows, the above correction will be taken
into account as an overall factor of 0.64 in front of the
branching ratios of the decays $\omega\to5\pi$. When making this
estimate, the imaginary part of the $\rho$ propagators in square
brackets of Eq.~(\ref{ampli2}) and (\ref{ampli3}) is neglected.
This assumption is valid with the accuracy better than $1\%$ in
width.

The evaluation of the partial widths valid with the accuracy
$20\%$ can be obtained upon using the expressions (\ref{ampli})-
(\ref{ampli4}). Such an accuracy is estimated by noting that the
numerical magnitude of the $\rho$ propagators in the expressions
for the $\omega\to5\pi$ decay amplitudes in Eqs.~(\ref{om1pi0})
and (\ref{om3pi0}) evaluated by assuming  the invariant mass of
the four pion system to be determined by either the "equilibrium"
pion energy or by the center of allowed range of the variation of
this mass, respectively, differs by the quantity not exceeding 10
$\%$ of the numerical value of the $\rho$ propagator. Defining the
branching ratio at the $\omega$ mass as
\begin{equation}
B_{\omega\to5\pi}=\Gamma_{\omega\to5\pi}/\Gamma_\omega,
\label{brom}
\end{equation}
one finds
\begin{eqnarray}
B_{\omega\to2\pi^+2\pi^-\pi^0}&=&\left|1-{D_\rho(\overline{m^2_{4\pi}})
\over2D_\rho(\overline{m^2_{2\pi}})}\right|^2\left({5\over2}\right)^2
{2\over\pi\Gamma_\omega}  \nonumber\\
&&\times\int_{4m_{\pi^+}}^{m_\omega-m_{\pi^0}}
dm   \nonumber\\
&&\times{m^2\Gamma_{\omega\to\rho^0\pi^0}(m)
\Gamma_{\rho\to2\pi^+2\pi^-}(m)\over|D_\rho(m^2)|^2}
\nonumber\\
&&=1.1\times10^{-9}
\label{b1pi0}
\end{eqnarray}
where $$\Gamma_{\omega\to\rho^0\pi^0}(m)=g^2_{\omega\rho\pi}q^3
(m_\omega,m,m_{\pi^0})/12\pi,$$
$$g_{\omega\rho\pi}={N_cg^2\over8\pi^2f_\pi}=14.3\mbox{
GeV}^{-1},$$ and the numerical data for the $\rho\to2\pi^+2\pi^-$
decay width obtained in Sec.~\ref{sec3} are used. Note also the
$a^{-1}$ dependence of the $\omega\to5\pi$ width on the HLS
parameter $a$. The branching ratio $B_{\omega\to\pi^+\pi^-3\pi^0}$
is obtained from Eq.~(\ref{b1pi0}) upon changing  the lower
integration limit to $m_{\pi^+}+3m_{\pi^0}$,  the substitution
$m_{\pi^0}\to m_{\pi^+}$ in the expression for the momentum $q$,
and the substitution of the $\rho^+\to\pi^+3\pi^0$ decay width
numerically calculated in Sec.~\ref{sec3}, instead of the
$\rho\to2\pi^+2\pi^-$ one. Note that the former is corrected for
the mass difference of charged and neutral pions. Of course, the
main correction of this sort comes from the phase space volume of
the final 4$\pi$ state. One obtains
\begin{eqnarray}
B_{\omega\to\pi^+\pi^-3\pi^0}&=&\left|1-{D_\rho(\overline{m^2_{4\pi}})
\over2D_\rho(\overline{m^2_{2\pi}})}\right|^2\left({5\over2}\right)^2
{2\over\pi\Gamma_\omega}  \nonumber\\
&&\times\int_{m_{\pi^+}+3m_{\pi^0}}^{m_\omega-m_{\pi^+}} dm
\nonumber\\ &&\times{m^2\Gamma_{\omega\to\rho^+\pi^-}(m)
\Gamma_{\rho^+\to\pi^+3\pi^0}(m)\over|D_\rho(m^2)|^2} \nonumber\\
&&=8.5\times10^{-10}, \label{b3pi0}
\end{eqnarray}
where $$\Gamma_{\omega\to\rho^+\pi^-}(m)=g^2_{\omega\rho\pi}q^3
(m_\omega,m,m_{\pi^+})/12\pi.$$ As is pointed out in
Ref.~\cite{bando}, the inclusion of the direct
$\omega\to\pi^+\pi^-\pi^0$ vertex reduces the 3$\pi$ decay width
of the $\omega$ by $33\%$. This implies that one should make the
following replacement to take into account the effect of  the
pointlike diagrams in Fig.~\ref{fig10}(b) in the expression for
the suppression factor:
\begin{eqnarray}
\left|1-{D_\rho(\overline{m^2_{4\pi}})
\over2D_\rho(\overline{m^2_{2\pi}})}\right|^2&\to&
\left|1-{D_\rho(\overline{m^2_{4\pi}})\over2}\left[{1\over
D_\rho(\overline{m^2_{2\pi}})}\right.\right.  \nonumber\\
&&\left.\left.-{1\over3m^2_\rho}\right]\right|^2
\approx\left|1-{D_\rho(\overline{m^2_{4\pi}})
\over3D_\rho(\overline{m^2_{2\pi}})}\right|^2    \nonumber\\
&&\simeq0.75,
\label{correc}
\end{eqnarray}
\narrowtext
instead of 0.64, which results in the increase of the above branching ratios
by the factor of 1.17.

\begin{figure}
\centerline {\epsfysize=7in \epsfbox{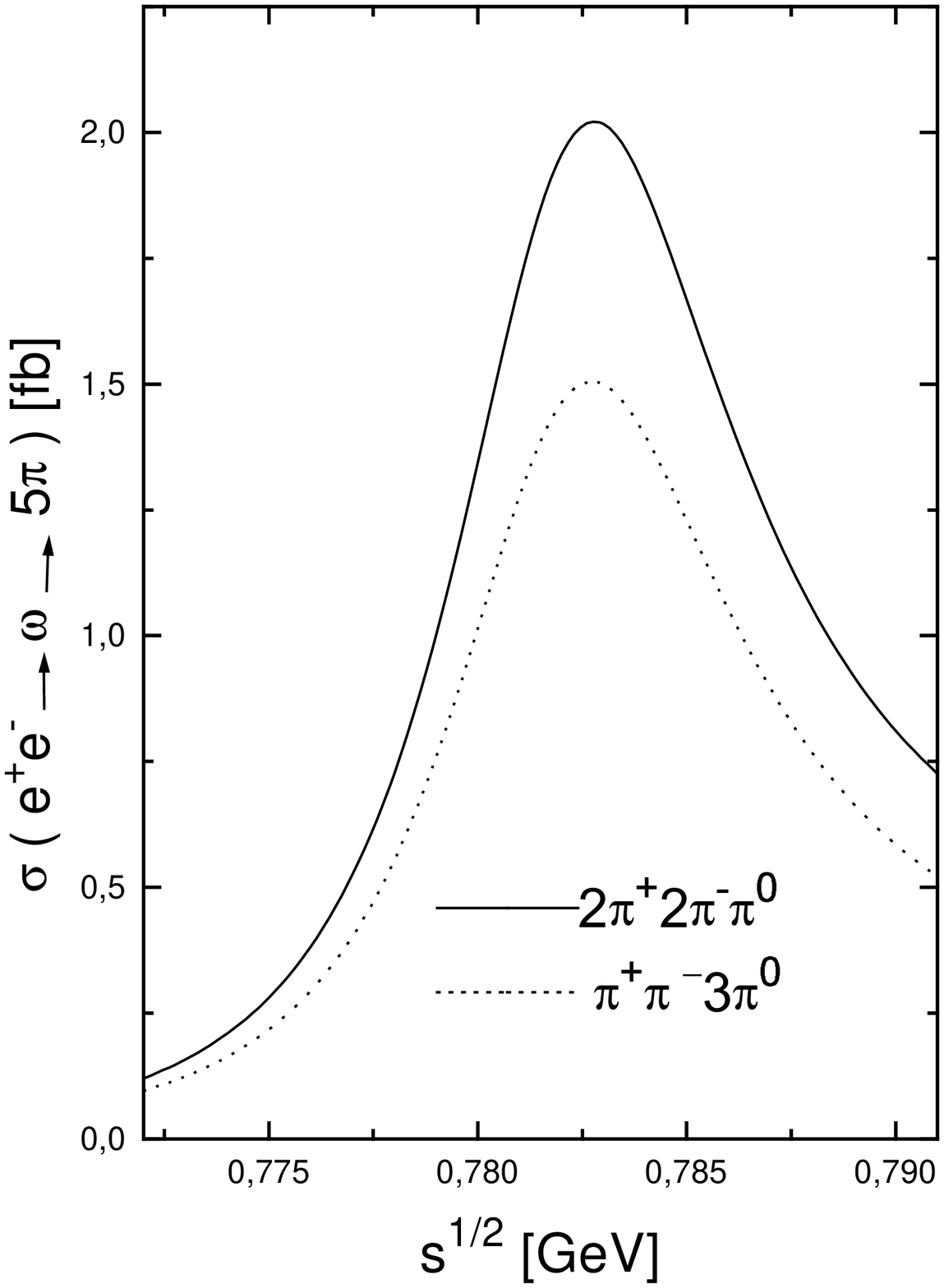}}
\caption{The $\omega\to5\pi$ excitation curves in
$e^+e^-$  annihilation in the vicinity of the $\omega$ resonance.
\label{fig11}}
\end{figure}

The numerical value of the $\omega\to5\pi$ decay width changes by
the factor of two when varying the energy within
$\pm\Gamma_\omega/2$ around the $\omega$ mass. In other words, the
dependence of this partial width on energy is very strong. This is
illustrated by Fig.~\ref{fig11} where the $\omega\to5\pi$
excitation curves in $e^+e^-$ annihilation,
\begin{eqnarray}
\sigma_{e^+e^-\to\omega\to5\pi}(s)&=&12\pi\left({m_\omega\over E}\right)^3
\Gamma_{\omega e^+e^-}(m_\omega)   \nonumber\\
&&\times{\Gamma_\omega B_{\omega\to5\pi}(E)\over\left[(s-m^2_\omega)^2+
(m_\omega\Gamma_\omega)^2\right]},
\label{omee}
\end{eqnarray}
are plotted. Here $B_{\omega\to2\pi^+2\pi^-\pi^0}(E)$
[$B_{\omega\to\pi^+\pi^-3\pi^0}(E)$] is given by Eq.~(\ref{b1pi0})
[(\ref{b3pi0})], respectively, with the substitution $m_\omega\to
E$. The mentioned strong energy dependence of the partial width
results in the asymmetric shape of the $\omega$ resonance and the
shift of its peak position by +0.7 MeV. As is seen from
Fig.~\ref{fig11}, the peak value of the $5\pi$ state production
cross section is about 1.5-2.0 femtobarns. Yet the decays
$\omega\to5\pi$ can be observable on $e^+e^-$ colliders. Indeed,
with  the luminosity $L=10^{33}\mbox{cm}^{-2}\mbox{s}^{-1}$ near
the $\omega$ peak, which seems to be feasible,  one may expect
about 2 events per week for the considered decays to be detected
at these colliders.

The angular distributions of the final pions  should be deduced
from the full amplitudes Eqs.~(\ref{om1pi0}) and (\ref{om3pi0}).
However, some qualitative conclusions about the angular
distributions can be drawn from the simplified expressions
Eqs.~(\ref{ampli}), (\ref{ampli2}), (\ref{ampli3}),
(\ref{ampli4}). Since helicity is conserved, only the states of
the $\omega(782)$ with the spin projections $\lambda=\pm1$ on the
$e^+e^-$ beam axes are populated. Corresponding expressions for
the various combination of the final pions can be found in
Appendix.

The strong energy dependence of the five pion partial width of the $\omega$
implies that the
branching ratio at the $\omega$ mass, Eq.~(\ref{brom}),
evaluated above, is slightly different from that determined by
the expression
\begin{equation}
B^{\rm aver}_{\omega\to5\pi}(E_1,E_2)=
{2\over\pi}\int_{E_1}^{E_2}dE{E^2\Gamma_\omega B_{\omega\to5\pi}(E)\over
(E^2-m^2_\omega)^2+(m_\omega\Gamma_\omega)^2}.
\label{bmom}
\end{equation}
Taking $E_1=$ 772 MeV and $E_2=$ 792 MeV, one finds $B^{\rm
aver}_{\omega\to2\pi^+2\pi^-\pi^0}(E_1,E_2)=9.0\times10^{-10}$ and
$B^{\rm
aver}_{\omega\to\pi^+\pi^-3\pi^0}(E_1,E_2)=6.7\times10^{-10}$ to
be compared to Eqs.~(\ref{b1pi0}) and (\ref{b3pi0}), respectively.
In particular, the quantity $B^{\rm
aver}_{\omega\to2\pi^+2\pi^-\pi^0}(E_1,E_2)$ is the relevant
characteristics  of this specific decay mode in photoproduction
experiments. The Jefferson Lab "photon factory" \cite{dzierba}
could also be suitable for detecting the five pion decays of the
$\omega$. However, in view of the suppression of the $\omega$
photoproduction cross section by the factor of 1/9 as compared
with the $\rho$ one, the total number of $\omega$ mesons will
amount to $7\times10^8$ per nucleon. Hence, the increase of
intensity of this machine by the factor of 50 is highly desirable,
in order to observe the decay $\omega\to5\pi$ and measure its
branching ratio. Evidently, the $\omega$ photoproduction on heavy
nuclei is preferable in view of the dependence of the cross
section on atomic weight $A$ growing as $A^{0.8-0.95}$
\cite{leith}.

The conclusions about how the angular distributions in the
$\omega$ photoproduction are related with those in $e^+e^-$
annihilation, are basically similar to those concerning the $\rho$
photoproduction discussed in Sec.~\ref{susecph}. The expressions
for these distributions coincide with the corresponding
expressions Eqs.~(\ref{anc1}), (\ref{anc2}), (\ref{ann1}), and
(\ref{ann2})  obtained  for the case of $e^+e^-$
annihilation. The vector ${\bf n}_0$ in those formulas can be
treated as pointed along the photon beam direction.

\section{Conclusion}
\label{sec5}

The results presented in this paper show that, in the
$\rho\to4\pi$ decay amplitude, the contributions of the higher
derivatives specified by the term induced by the anomalous
Lagrangian of Wess and Zumino, vanish rapidly when decreasing the
invariant mass of the four pion system below 700 MeV. The loop
corrections are also expected to behave similarly.  The decay
$\omega\to5\pi$ is of a special interest, because its kinematics
is such that the final pions are essentially nonrelativistic, and
the above effects are completely suppressed in its decay
amplitude. Hence the approach to the decays of the vector mesons
$\rho$ and $\omega$  presented in this paper is the  zero order
approximation to the full amplitude, in a close analogy with the
Weinberg amplitude in the classical $\pi\pi$ scattering. Under the
approximation of the present paper, all chiral models of the
vector meson interactions with pions \cite{fn1} are
indistinguishable in their predictions concerning the many pion
branching ratios. Any difference could manifest upon including the
higher derivatives and chiral loops. As the next step in the
development of the present study, the inclusion of the $a_1$ meson
contribution at the tree level approximation would be important.
This could help in extending the validity of the treatment up to
the invariant masses  just below 910 MeV. The task looks
meaningful, since the loop corrections, whose particular
manifestation is the finite width effects uncovered to be
unimportant at $m_{4\pi}<910$ MeV ( see Sec.~\ref{sec3}), are
still expected to be small at these invariant masses.

In our opinion, the above considerations show that the left
shoulder of the $\rho$ peak is the very perspective place to study
the effects of chiral dynamics of the vector meson interactions.
The $e^+e^-$ colliders with the large enough luminosity at
energies below the $\rho$ mass could provide the controlled source
of soft pions. The role of  higher derivatives, loop corrections,
and, possibly, the $\rho^\prime$, $\rho^{\prime\prime}$
contributions in the low energy effective Lagrangian for the soft
pions, as well as various schemes of incorporation of the vector
mesons into the chiral approach, can be successfully tested with
such machines. The intense beams of photons from the Jefferson
Laboratory "photon factory" are also of great importance in
achieving the mentioned theoretical goals. Certainly, the
measurements of the branching ratio of the five pion decays of the
$\omega$ at the level of $B_{\omega\to5\pi}\sim 10^{-9}$
constitute the real challenge to experimenters, but by the reasons
specified above the effects of chiral dynamics of the vector meson
interactions are manifested in  the decay $\omega\to5\pi$ in the
very  clean way.

\acknowledgements
We are grateful to G.~N.~Shestakov and A.~M.~Zaitsev for
discussion. The present work is supported in part by
grant No. RFBR-INTAS IR-97-232.

\appendix
\section*{Angular distributions}
\label{app}

Here a number of expressions for the angular distributions of
various combinations of the final pions in the decays
$\rho\to4\pi$ and $\omega\to5\pi$ are given. In what follows, the
one photon  $e^+e^-$ annihilation production mechanism for these
states is assumed, where the $e^+e^-$ beam axes is characterized
by the unit vector ${\bf n}_0$ directed along the z axes.

\subsection{The angular distributions in the $\rho\to4\pi$ decay}
\label{apprho}

Taking $\theta_i,\phi_i$ to be the polar and azimuthal angles of
the pion three momentum ${\bf q}_i$, where the momentum assignment
corresponds to Eq.~(\ref{nonrel}), one finds the following.

(i) The $\rho^0\to2\pi^+2\pi^-$ decay. The probability density of
the emission of four charged pions can be found directly from the
first Eq.~(\ref{nonrel}): \widetext
\begin{eqnarray}
w&\propto&({\bf q}_1+{\bf q}_2-{\bf q}_3-{\bf q}_4)^2 -[{\bf
n}_0({\bf q}_1+{\bf q}_2-{\bf q}_3-{\bf q}_4)]^2\nonumber\\
&&=\sum_{i=1}^4{\bf q}_i^2\sin^2\theta_i +2|{\bf
q}_1|(1-P_{23}-P_{24})|{\bf q}_2|\sin\theta_1\sin\theta_2
\cos(\phi_1-\phi_2)\nonumber\\ &&-2|{\bf q}_2|(1+P_{34})|{\bf
q}_3|\sin\theta_2\sin\theta_3 \cos(\phi_2-\phi_3)\nonumber\\
\nonumber\\ &&+2|{\bf q}_3||{\bf q}_4|\sin\theta_3\sin\theta_4
\cos(\phi_3-\phi_4). \label{4pipm}
\end{eqnarray}
\narrowtext
 One may use the relation
\begin{equation}
(\varepsilon,q_1+q_2+q_3+q_4)=0 \label{transv}
\end{equation}
that expresses the transverse character of the $\rho$ polarization
four vector $\varepsilon$, to get rid of the momenta of negatively
charged pions $q_3$ and $q_4$. Then the probability density of the
emission of two $\pi^+$'s  found from the first Eq.~(\ref{nonrel})
is
\begin{eqnarray}
w&\propto&({\bf q}_1+{\bf q}_2)^2 -[{\bf n}_0({\bf q}_1+{\bf
q}_2)]^2\nonumber\\ &&={\bf q}_1^2\sin^2\theta_1+{\bf
q}_2^2\sin^2\theta_2 +2|{\bf q}_1||{\bf
q}_2|\sin\theta_1\sin\theta_2\nonumber\\
&&\times\cos(\phi_1-\phi_2). \label{2pipl}
\end{eqnarray}
Allowing for Eq.~(\ref{transv}), the angular distribution for the
emission of two $\pi^-$'s is obtained from Eq.~(\ref{2pipl}) upon
the replacement ${\bf q}_{1,2}\to{\bf q}_{3,4}$.

(ii) The $\rho^0\to\pi^+\pi^-2\pi^0$ decay. The probability density
of the emission of  $\pi^+\pi^-$ pair  found from the second
Eq.~(\ref{nonrel}) in the form
\begin{eqnarray}
w&\propto&({\bf q}_1-{\bf q}_2)^2 -[{\bf n}_0({\bf q}_1-{\bf
q}_2)]^2\nonumber\\ &&={\bf q}_1^2\sin^2\theta_1+{\bf
q}_2^2\sin^2\theta_2 -2|{\bf q}_1||{\bf
q}_2|\sin\theta_1\sin\theta_2\nonumber\\
&&\times\cos(\phi_1-\phi_2). \label{piplmin}
\end{eqnarray}
Getting rid of the momentum $q_2$ one finds the corresponding
expression for the final state $\pi^+2\pi^0$: \widetext
\begin{eqnarray}
w&\propto&(2{\bf q}_1-{\bf q}_3-{\bf q}_4)^2 -[{\bf n}_0(2{\bf
q}_1-{\bf q}_3-{\bf q}_4)]^2\nonumber\\ &&=4{\bf
q}_1^2\sin^2\theta_1+{\bf q}_3^2\sin^2\theta_3 +{\bf
q}_4^2\sin^2\theta_4 -(1+P_{34})4|{\bf q}_1||{\bf
q}_3|\sin\theta_1\sin\theta_3 \cos(\phi_1-\phi_3)\nonumber\\
&&+2|{\bf q}_3||{\bf q}_4|\sin\theta_3\sin\theta_4
\cos(\phi_3-\phi_4), \label{pipl2pi0}
\end{eqnarray}
\narrowtext where $P_{ij}$ interchanges the pion momenta $q_i$ and
$q_j$. In view of Eq.~(\ref{transv}), the angular distribution for
the state $\pi^-2\pi^0$ is obtained from the above upon the
replacement ${\bf q}_1\to{\bf q}_2$ and changing the signs in
front of the terms containing $(1+P_{34})$.

\subsection{The angular distributions in the $\omega\to5\pi$ decay}
\label{appom}

In what follows the suitable notation for the vector product of
the pion momenta are used:
\begin{eqnarray}
[{\bf q}_i\times{\bf q}_j]&=&|{\bf q}_i||{\bf q}_j|\sin\theta_{ij}
\nonumber\\
&&\times(\sin\Theta_{ij}\cos\Phi_{ij},\sin\Theta_{ij}\sin\Phi_{ij},
\cos\Theta_{ij}). \label{not}
\end{eqnarray}
In other words, $\theta_{ij}$ is the angle between the pion
momenta ${\bf q}_i$ and ${\bf q}_j$, $\Theta_{ij}$, $\Phi_{ij}$
being the polar and azimuthal angles of the normal to the plane
spanned by the  momenta ${\bf q}_i$ and ${\bf q}_j$. Choosing
${\bf n}_0$ to be the unit vector along z axes, the probability
density of the emission of two $\pi^+$'s with the momenta ${\bf
q}_1$, ${\bf q}_2$, and $\pi^0$ with the momentum ${\bf q}_4$ is
represented as
\begin{eqnarray}
w&\propto&\left[{\bf q}_4\times\left({\bf q}_1+{\bf
q}_2\right)\right]^2-\left({\bf n}_0\cdot\left[{\bf
q}_4\times\left({\bf q}_1+{\bf q}_2\right)\right]\right)^2
\nonumber\\ &&={\bf q}^2_4\left[{\bf
q}^2_1\sin^2\theta_{41}\sin^2\Theta_{41}+ {\bf
q}^2_2\sin^2\theta_{42}\sin^2\Theta_{42}\right. \nonumber\\
&&\left.+2|{\bf q}_1||{\bf q}_2|\sin\Theta_{41}\sin\Theta_{42}
\sin\theta_{41}\sin\theta_{42}\right.\nonumber\\
&&\left.\times\cos(\Phi_{41}-\Phi_{42})\right] \label{anc1}
\end{eqnarray}
in the case of the final state $2\pi^+2\pi^-\pi^0$. Here the
momentum assignment is the same as in Eq.~(\ref{om1pi0}). The
angular distribution of two $\pi^-$'s with the momenta ${\bf
q}_3$, ${\bf q}_5$, and $\pi^0$ is obtained from Eq.~(\ref{anc1})
upon the replacement ${\bf q}_{1,2}\to{\bf q}_{3,5}$, because the
identity
$$\varepsilon_{\mu\nu\lambda\sigma}q_\mu\epsilon_\nu(q_1+q_2)_\lambda
q_{4\sigma}=-\varepsilon_{\mu\nu\lambda\sigma}q_\mu\epsilon_\nu
(q_3+q_5)_\lambda q_{4\sigma}$$ is valid. Since another identity
$$\varepsilon_{\mu\nu\lambda\sigma}q_\mu\epsilon_\nu(q_1+q_2)_\lambda
q_{4\sigma}=-\varepsilon_{\mu\nu\lambda\sigma}q_\mu\epsilon_\nu
(q_1+q_2)_\lambda(q_3+q_5)_\sigma$$ is valid, one can write the
angular distribution that includes four charged pions: \widetext
\begin{eqnarray}
w&\propto&[({\bf q}_1+{\bf q}_2)\times({\bf q}_3+{\bf q}_5)]^2
-\left({\bf n}_0\cdot [({\bf q}_1+{\bf q}_2)\times({\bf q}_3+{\bf
q}_5)]\right)^2 \nonumber\\ &&=(1+P_{12})(1+P_{35}){\bf q}^2_1{\bf
q}^2_3 \sin^2\theta_{13}\sin^2\Theta_{13}\nonumber\\ &&+2|{\bf
q}_1||{\bf q}_2|(1+P_{35}){\bf
q}_3^2\sin\theta_{13}\sin\theta_{23}
\sin\Theta_{13}\sin\Theta_{23}\cos(\Phi_{13}-\Phi_{23})
\nonumber\\ &&+2|{\bf q}_3||{\bf q}_5|(1+P_{12}){\bf
q}_1^2\sin\theta_{13}\sin\theta_{15}
\sin\Theta_{13}\sin\Theta_{15}\cos(\Phi_{13}-\Phi_{15})
\nonumber\\ &&+2|{\bf q}_1||{\bf q}_2||{\bf q}_3||{\bf
q}_5|(1+P_{35})
\sin\theta_{13}\sin\theta_{25}\sin\Theta_{13}\sin\Theta_{25}
\cos(\Phi_{13}-\Phi_{25}). \label{anc2}
\end{eqnarray}
\narrowtext Here $P_{ij}$  interchanges the indices $i$ and $j$.
In the case of the final state $\pi^+\pi^-3\pi^0$ the
corresponding probability density can be obtained from
Eqs.~(\ref{ampli3}) and (\ref{ampli4}) and looks as
\begin{eqnarray}
w&\propto&[{\bf q}_1\times{\bf q}_2]^2-({\bf n}_0\cdot[{\bf
q}_1\times{\bf q}_2])^2       \nonumber\\ &&={\bf q}^2_1{\bf
q}^2_2\sin^2\theta_{21}\sin^2\Theta_{21}. \label{ann1}
\end{eqnarray}
Here the momentum assignment is the same as in Eq.~(\ref{om3pi0}).
The corresponding angular distribution of one charged, say
$\pi^+$, and three neutral pions can be obtained from
Eqs.~(\ref{ampli3}) and (\ref{ampli4}) upon using the identity
$$\varepsilon_{\mu\nu\lambda\sigma}q_\mu\epsilon_\nu
q_{1\lambda}q_{2\sigma}
=-\varepsilon_{\mu\nu\lambda\sigma}q_\mu\epsilon_\nu
q_{1\lambda}(q_3+q_4+q_5)_\sigma$$ and looks as
\begin{eqnarray}
w&\propto&\left[{\bf q}_1\times\sum_i{\bf q}_i\right]^2
-\left({\bf n}_0\cdot\left[{\bf q}_1\times\sum_i{\bf
q}_i\right]\right)^2 \nonumber\\ &&={\bf q}^2_1\left[\sum_i{\bf
q}^2_i\sin^2\theta_{i1}\sin^2\Theta_{i1} \right. \nonumber\\
&&\left.+2\sum_{i\not=j}|{\bf q}_i||{\bf
q}_j|\sin\theta_{i1}\sin\theta_{j1}
\sin\Theta_{i1}\sin\Theta_{j1}\right.\nonumber\\
&&\left.\times\cos(\Phi_{i1}-\Phi_{j1})\right]. \label{ann2}
\end{eqnarray}
Here indices $i,j$ run over 3,4,5.


\begin{references}
\bibitem{fn2}
Taking the higher derivatives into account demands also
taking into account the chiral loops, the task which is not yet fulfilled
for vector mesons.
\end{references}
\end{document}